\documentclass[12pt]{article}
\usepackage{epsfig}
\usepackage{amssymb}
\usepackage{pslatex}

\makeatletter

\@addtoreset{equation}{section}
\makeatother
\newcommand{\bea}{\begin{eqnarray}}
\newcommand{\eea}{\end{eqnarray}}
\newcommand{\be}{\begin{eqnarray}}
\newcommand{\ee}{\end{eqnarray}}

\def\tr{\mathop{\rm Tr}}

\def\fR{{\mathfrak R}}

\def\rmd{{\rm d}}

\begin{document}

\begin{titlepage}
\vskip-0cm
\begin{flushright}
SNUST 080701
\end{flushright}
\centerline{\Large \bf  Integrable Spin Chain}
\vskip0.25cm
\centerline{\Large \bf in}
 \vskip0.25cm
 \centerline{\Large \bf Superconformal Chern-Simons Theory }
\vspace{0.75cm}
\centerline{\large Dongsu Bak$^a$, \,\,\, Soo-Jong Rey$^b$}
\vspace{0.75cm}
\centerline{\sl Physics Department, University of Seoul, Seoul 130-743 {\rm KOREA}}
\vskip0.25cm
\centerline{\sl School of Physics \& Astronomy, Seoul National University, Seoul 151-747 {\rm KOREA}}
\vskip0.25cm
\centerline{\tt dsbak@uos.ac.kr \,\,\,\, sjrey@snu.ac.kr}
\vspace{1.0cm}
\centerline{ABSTRACT}
\vspace{0.75cm}
\noindent
${\cal N} = 6$ superconformal Chern-Simons theory was proposed as gauge theory dual to Type IIA string theory on AdS$_4 \otimes \mathbb{CP}^3$. We study integrability of the theory from conformal dimension spectrum of single trace operators at planar limit. At strong `t Hooft coupling, the spectrum is obtained from excitation energy of free superstring on OSp$(6|2,2;\mathbb{R})/$SO$(3,1)\times $SU$(3)\times$U(1) supercoset. We recall that the worldsheet theory is integrable classically by utilizing well-known results concerning sigma model on symmetric space. With R-symmetry group SU(4), we also solve relevant Yang-Baxter equation for a spin chain system associated with the single trace operators. From the solution, we construct alternating spin chain Hamiltonian involving three-site interactions between ${\bf 4}$ and $\overline{\bf 4}$. At weak `t Hooft coupling, we study gauge theory perturbatively, and calculate action of dilatation operator to single trace operators up to two loops. To ensure consistency, we computed all relevant Feynman diagrams contributing to the dilatation opeator. We find that resulting spin chain  Hamiltonian matches with the Hamiltonian derived from Yang-Baxter equation. We further study new issues arising from the shortest gauge invariant operators Tr$Y^I Y^\dagger_J = ({\bf 15}, {\bf 1})$.  We observe that `wrapping interactions' are present, compute the true spectrum and find that
the spectrum agrees with prediction from supersymmetry. We also find that scaling dimension computed naively from alternating spin chain Hamiltonian coincides with the true spectrum. We solve Bethe ansatz equations for small number of excitations, and find indications of correlation between excitations of ${\bf 4}$'s and $\overline{\bf 4}$'s and of nonexistence of mesonic $({\bf 4} \overline{\bf 4})$ bound-state.

\end{titlepage}

\section{Introduction}
In a recent remarkable development, Aharony, Bergman, Jeffries and Maldacena (ABJM) \cite{Aharony:2008ug} made a new addition to the list of microscopic AdS/CFT correspondence \cite{Maldacena:1997re}: three-dimensional ${\cal N}=6$ superconformal Chern-Simons theory dual to Type IIA string theory on AdS$_4 \times \mathbb{CP}^3$ \cite{Nilsson:1984bj}. Both sides of the correspondence are characterized by two integer-valued coupling parameters $N$ and $k$. On the superconformal Chern-Simons theory side, they are the rank of product gauge group U($N) \times \overline{{\rm U}(N)}$ and Chern-Simons levels $+k, -k$, respectively. On the Type IIA string theory side, they are related to spacetime curvature and dilaton gradient or Ramond-Ramond flux, all measured in string unit. Much the same way as the counterpart between ${\cal N}=4$ super Yang-Mills theory and Type IIB string theory on AdS$_5 \times \mathbb{S}^5$, we can put the new correspondence into precision tests in the planar limit:
\bea
N \rightarrow \infty,  \qquad k \rightarrow \infty \qquad
\mbox{with} \qquad \lambda \equiv {N \over k} \quad \mbox{fixed}
\eea
by interpolating `t Hooft coupling parameter $\lambda$ between superconformal Chern-Simons theory regime at $\lambda \ll 1$ and semiclassical AdS$_4 \times \mathbb{CP}^3$ string theory regime at $\lambda \gg 1$.

In the correspondence between ${\cal N}=4$ super Yang-Mills theory and Type IIB string theory on AdS$_5 \times \mathbb{S}^5$, the integrability
structure first discovered by Minahan and Zarembo \cite{Minahan:2002ve} led to remarkable progress
in diverse fronts of the correspondence~\footnote{Selected but nonexhaustive list of contributions in this subject include \cite{Beisert:2003tq} - \cite{Beisert:2006ez}. For a comprehensive mid-development review, see \cite{Beisert:2004ry}.}. It is therefore interesting to examine if the new correspondence shows also an integrability structure. The purpose of this work is to demonstrate integrability structure inherent to the ${\cal N}=6$ superconformal Chern-Simons theory of ABJM~\footnote{Integrability in ${\cal N} \le 3$ superconformal Chern-Simons theory was investigated by Gaiotto and Yin~\cite{Gaiotto:2007qi} previously. Their tentative result indicated otherwise.}.

AdS/CFT correspondence asserts that gauge invariant, single trace operators in superconformal Chern-Simons theory are dual to free string excitation modes in AdS$_4 \times \mathbb{CP}^3$, valid at weak and strong `t Hooft coupling regime, respectively. In particular, conformal dimension of the operators should match with excitation energy of the string modes. The ${\cal N}=6$ superconformal Chern-Simons theory has SO(6)$\simeq$SU(4) R-symmetry and contains two sets of bi-fundamental scalar fields $Y^I, Y^\dagger_I$ $(I=1,2,3,4)$ that transform as ${\bf 4}, \overline{\bf 4}$ under SU(4). Therefore, the single trace operators take the form:
\bea
{\bf \cal O} &=& \mbox{Tr} (Y^{I_1} Y^\dagger_{J_1} \cdots Y^{I_L} Y^\dagger_{J_L} ) C^{J_1 \cdots J_L}_{I_1 \cdots I_L} \nonumber \\
&=& \overline{\mbox{Tr}} (Y^\dagger_{J_1} Y^{I_1} \cdots Y^\dagger_{J_L} Y^{I_L}) C^{J_1 \cdots J_L}_{I_1 \cdots I_L} \ . \label{STO}
\eea
In superconformal Chern-Simons theory, chiral primary operators, corresponding to the choice of (\ref{STO}) with $C^{J_1 \cdots J_L}_{I_1 \cdots I_L}$ totally symmetric in both sets of indices and traceless, form the lightest states. In free string theory on AdS$_4 \times \mathbb{CP}^3$. Kaluza-Klein supergravity modes form the lightest states. In this work, we study conformal dimension of single trace operators and identify integrability structure organizing the excitation spectrum above the chiral primary or the Kaluza-Klein states.

In section 2, we begin with recapitulating the standard argument for integrability of free string on AdS$_4 \times \mathbb{CP}^3$ at $\lambda \rightarrow \infty$. Recalling the construction of~\cite{Brezin:1979am} and utilizing the idea of~\cite{Bena:2003wd}, we argue that sigma model on OSp$(6 \vert 2, 2, \mathbb{R})/$[SO$(3,1)\times$SU$(3)\times$U(1)] supercoset has commuting monodromy matrices and infinitely many conserved nonlocal charges.
In section 3, we begin main part of this work. Guided by earlier development in ${\cal N}=4$ super Yang-Mills counterpart, we assume
integrability and solve Yang-Baxter equations for R-matrices between ${\bf 4}$ and $\overline{\bf 4}$ sites in (\ref{STO}). From corresponding transfer matrices, we then find the Hamiltonian takes
the form of one-parameter family of 'alternating spin chain', whose variants were studied previously in different contexts~\cite{deVega:1991rc}-\cite{Ragoucy:2007kg}. In section 4, we study superconformal Chern-Simons theory of
ABJM at $\lambda \rightarrow 0$ in perturbation theory. Pure Chern-Simons theory is
free from any ultraviolet divergences since the theory is diffeomorphism invariant and hence topological . Once matter is coupled, as in ABJM theory, topological feature is lost and the quantum theory will receive nontrivial radiative corrections. As such, the single trace operators (\ref{STO}) will acquire nontrivial anomalous dimensions in general. In three dimensions, logarithmic ultraviolet divergence arises only at even loop orders. Therefore, the first nontrivial correction starts at two loops. We compute two-loop operator mixing and anomalous dimension matrix of the single trace operators (\ref{STO}). In dimensional reduction method, we compute the complete set of relevant Feynman diagrams and find that the two-loop anomalous dimension matrix matches with the integrable `alternating spin chain' Hamiltonian derived in section 3.
In section 5, we study a new important feature of the superconformal Chern-Simons theory compared to ${\cal N}=4$ super Yang-Mills theory. Since the anomalous dimensions begin to arise from two loops and next-to-nearest sites, the shortest single trace operators of $L=1$ will be subject to `wrapping interactions'. The `alternating spin chain' Hamiltonian does not describe spectrum of
$L=1$ operators, so we compute all relevant `wrapping interaction' diagrams and construct the correct
Hamiltonian for $L=1$. Curiously, we find that the correct spectrum coincides with the naive spectrum
computed from the `alternating spin chain' Hamiltonian at $L=1$. In section 6, utilizing results previously obtained for general $A_{n-1}$ Lie algebras \cite{Kulish:1983rd, Aladim:1993uu, Ragoucy:2007kg}, we explicitly write down eigenvalues of the transfer matrices and Bethe ansatz equations of the `alternating spin chain' we derived in section 3. To gain understanding how the `alternating spins' behave, we solve the equations for a few simple situations. We find an indication for real-space correlations between excitations on ${\bf 4}$ spin sites and those on $\overline{\bf 4}$ spin sites, and for non-existence of meson-like $({\bf 4} \overline{\bf 4}$) bound-states. We discuss various implications of these findings for general excitations. In particular, we argue that general excitations are more complex than the pattern emerging from closed SU(2) sub-sectors discussed recently \cite{Gaiotto:2008cg, Grignani:2008is}.

$\bullet$ \underline{Note added}: While bulk of this work was completed, we received the preprint by Minahan and Zarembo \cite{Minahan:2008hf}, which deals with issues overlapping with ours.
Version 1 of their preprint was based on ungrounded Feynman rules and incomplete set of contributing Feynman diagrams. In light of utmost importance of precise evaluation, we decided to carefully carry out various internal consistency checks before releasing our results. Meanwhile, we also received preprints by Arutyunov and Frolov \cite{Arutyunov:2008if}, which overlaps with section 2 and substantiate several pertinent issues including kappa symmetry. We also note the preprints by Stefanski \cite{Stefanski:2008ik} and by Gromov and Vieira \cite{Gromov:2008bz} on the same issue.

\section{Integrable String from Worldsheet Sigma Model}
In this section, we set out a motivation for searching for integrability in ${\cal N}=6$ superconformal Chern-Simons theory. The $\lambda \rightarrow \infty$ dual of this theory is Type IIA string theory on AdS$_4 \otimes \mathbb{CP}^3$. The background is a direct product of symmetric spaces, AdS$_4$ and $\mathbb{CP}^3$. It is well known that the $(1+1)$-dimensional sigma model on symmetric space is classically integrable. So, at least for bosonic modes, we expect worldsheet dynamics of a free string on AdS$_4 \otimes \mathbb{CP}^3$ is integrable at the classical level, $\lambda \rightarrow \infty$. In this section, we recapitulate this argument for the bosonic
part and discuss how the construction to full superstring can be made.

Bosonic part of string worldsheet Lagrangian on AdS$_4 \otimes \mathbb{CP}^3$ is given by
\bea
I_{\rm b} = {R^2 \over 4 \pi} \int_\Sigma \sqrt{-h} h^{\alpha \beta} \Big[
(D_\alpha X^m)^\dagger (D_\beta X^m) + (D_\alpha Z^a)^\dagger (D_\beta Z^a) \Big].
\label{waction}
\eea
Here, we use embedding coordinates in $\mathbb{R}^{3,2}$ and $\mathbb{C}^{4}$ and describe
AdS$_4 =$SO(3,2)/SO(3,1) and $\mathbb{CP}^3=$SU(4)/SU(3)$\times$U(1) as $G/H$ coset hypersurfaces:
\bea
\mbox{AdS}_4 :&& \qquad (X^m) = (X^{-1}, X^1, X^2, X^3, X^0) \hskip1.2cm \mbox{with} \qquad X^2 = 1 \, ,  \nonumber \\
\mathbb{CP}^3 :&& \qquad (Z^a) = (Z^1, Z^2, Z^3, Z^4)/{\{\simeq, \mathbb{C}\}} \qquad \mbox{with} \qquad
|Z|^2 = 1,
\eea
respectively. The hypersurface conditions are
imposed by introducing auxiliary connection $K_\alpha, A_\alpha$ and by defining covariant derivatives~\footnote{We introduced auxiliary connection $K_\alpha$ to treat AdS$_4$ in complete
parallel to $\mathbb{CP}^3$.}
$D_\alpha X^m \equiv \partial_\alpha X^m + i K_\alpha X^m$ and $D_\alpha Z^a \equiv \partial_\alpha Z^a + i A_\alpha Z^a$. These conditions imply that
$X^m \partial_\alpha X_m = 0$ and $(D_\alpha Z^a)^\dagger Z^a = Z^{a\dagger} (D_\alpha Z^a) = 0$.
Following~\cite{Brezin:1979am}, we first recapitulate basic aspects for classical integrability of sigma model on AdS$_4 \times \mathbb{CP}^3$. Construction of the coset sigma model is facilitated by the coset elements:
\bea
G(\sigma) \equiv g(\sigma)\oplus\widetilde{g}(\sigma) = e^{i \pi P(\sigma)}\oplus e^{i \pi \widetilde{P}(\sigma)}
\label{G}\eea
where $P(\sigma), \widetilde{P}(\sigma)$ are projection matrices onto respective one-dimensional subspaces. They are
\bea
P^{mn}(\sigma) &=& X^m(\sigma) X^n(\sigma) \qquad \mbox{with} \qquad \delta_{mn} P^{mn}(\sigma) = 1, \nonumber \\
\widetilde{P}^{ab} (\sigma) &=& {Z^a}^\dagger (\sigma) Z^b (\sigma) \qquad \mbox{with} \qquad  \delta_{ab} P^{ab} (\sigma) = 1, \label{P}
\eea
respectively. By elementary algebra, we verify that
\bea
G(\sigma)  = G^{-1}(\sigma) = (\mathbb{I}_5 - 2 P(\sigma))\oplus(\mathbb{I}_4 - 2 \widetilde{P}(\sigma)).
\label{G2}\eea
Then, because $- 8 |D_\alpha X^m|^2 = {\rm Tr} (\partial_\alpha g \cdot \partial_\alpha g^{-1})$ for AdS$_4$ and $+ 4 |D_\alpha Z^a|^2 = {\rm Tr} (\partial_\alpha \widetilde{g} \cdot \partial_\beta \widetilde{g}^{-1})$ for $\mathbb{CP}^3$, the worldsheet action (\ref{waction}) is expressible as
\bea
I_{\rm bosonic} = {R^2 \over 8} \int_\Sigma \sqrt{-h} h^{\alpha \beta} \Big[ -{1 \over 2} \mbox{Tr} J_\alpha J_\beta + 2 \mbox{Tr} \widetilde{J}_\alpha \widetilde{J}_\beta \Big],
\label{b-wsaction}
\eea
where $J = g^{-1} {\rm d} g$ and $\tilde{J} = \tilde{g}^{-1} {\rm d} \tilde{g}$, respectively.
We shall choose the conformal gauge $\sqrt{-h} h^{\alpha \beta} = \delta^{\alpha \beta}$ on the worldsheet. This leads to Virasoro gauge condition
\bea
T_\pm \equiv -{1 \over 4} ( J_0 \pm J_1)^2 + (\tilde{J}_0 \pm \tilde{J}_1)^2 = 0 \ .
\eea
The currents $J, \tilde{J}$ are conserved by equations of motion, and define tangent flows on the
$G/H$ coset space.

We now take group conjugation and transform the left-invariant currents $J, \widetilde{J}$ to the right-invariant currents:
$(j, \widetilde{j}) = (g \cdot J \cdot g^{-1}, \,\, \widetilde{g} \cdot \widetilde{J} \cdot \tilde{g}^{-1}).$
The equations of motion in conformal gauge are
\be
{\rm d} {}^* j(\sigma) = 0 \qquad \mbox{and} \qquad {\rm d} {}^* \widetilde{j}(\sigma) = 0.
\label{eom}
\ee
From the Bianchi identities, we also have
\bea
d j +  j \wedge j = 0 \qquad \mbox{and} \qquad d \widetilde{j} + \widetilde{j} \wedge \widetilde{j} = 0.
\label{bianchi}
\eea
Finally, Virasoro constraints are
\bea
-{1 \over 4} (j_0 \pm j_1)^2 + (\tilde{j}_0 \pm \widetilde{j}_1)^2 = 0.
\eea

We can solve these equations using the Lax representation. Consider the Lax derivative with flat connection $a(x)$ depending on a spectral parameter $x$:
\bea
D(x) = {\rm d} + a(x) \qquad \mbox{with} \qquad {\rm d} a + a \wedge a = 0.
\eea
Using (\ref{eom}, \ref{bianchi}), we find that the most general form of the Lax connection is
given by
\bea
a(x) = {2 \over x^2 - 1} j(\sigma) +{2 x \over x^2 -1} {}^* j(\sigma) \qquad \quad
x \in \mathbb{C} + \{\infty\}/\{\pm 1\}.
\eea
and similarly construct $\widetilde{a}(x)$ from $\tilde{j}(\sigma)$. With the flat connection
$A(x) \equiv (a(x), \widetilde{a}(x))$, consider the Wilson line
\bea
W[\gamma; x] = {\cal P} \exp \Big( \int_\gamma A(x) \Big)
\eea
As the connection $A(x)$ is flat, the eigenvalues of the Wilson line are independent of the
choice of the contour $\gamma$. Thus, all Wilson lines commute each another and provides classical R-matrices obeying Yang-Baxter equations. Expanding
in spectral parameter $x$, we then obtain infinitely many conserved nonlocal charges
as moment of the power series:
\bea
{\cal Q}^{(n)} = {1 \over n!} \partial_x^n \int_\gamma {\rm d} \sigma A(x) \Big\vert_{x=0}.
\eea
This establishes that the sigma model on AdS$_4 \times \mathbb{CP}^3$ is classically integrable.

We now discuss how the above consideration may be extended to Type IIA string on the supercoset:
\bea
{\widehat{G} \over H} = {{\rm OSp}(6 \vert 2, 2, \mathbb{R}) \over {\rm SO}(3,1) \times {\rm SU}(3) \times {\rm U}(1)}.
\eea
With the coefficients of current bilinears determined as in (\ref{b-wsaction}), we see immediately
that the worldsheet Lagrangian is expressible as supertrace over the supergroup OSp$(6 \vert 2, 2, \mathbb{R})$:
\bea
I_{\rm supercoset} = {R^2 \over 8} \int_\Sigma \mbox{Str} \, (\widehat{J} \wedge {}^* \widehat{J}) \, .
\eea
Here, $\widehat{J}(\sigma) = \widehat{G}^{-1}(\sigma) {\rm d} \widehat{G}(\sigma)$ and $\widehat{G} (\sigma) = \exp ( i \pi \widehat{P}(\sigma))$ is the supercoset element. This indicates that the bosonic action (\ref{b-wsaction}) is extendible straightforwardly to a supercoset action by adding 24 fermionic off-diagonal components to (\ref{G}-\ref{G2}) and define super-projection matrix $\widehat{P}$ and supercoset element $\widehat{G}$ analogously. 

Construction of infinitely many nonlocal currents requires a new condition to the supercoset. If the supergroup $\hat{G}$ permits $\mathbb{Z}_4$ grading under which the subgroup $H$ is a fixed point set, the construction of~\cite{Bena:2003wd} implies that a flat connection exists from which nonlocal currents can be constructed through the Lax formulation. From the embedding we constructed of, we have $\widehat{J} = J + Q$, where $Q$ denotes fermionic current. For the supergroup we deal with, $\widehat{G} = $ OSp$(6 \vert 2, 2)$, it is well known that $\widehat{G}$ admits no outer automorphism of order four \cite{osp-auto}. However, one can easily construct a suitable $\mathbb{Z}_4$ inner automorphism. Since we need the subgroup $H$ is a fixed point set, the automorphism can be defined as a product of two $\mathbb{Z}_2$ involutions on the defining representations of SU(4)$\simeq$SO(6) and Sp(4) modulo overall reflection. This then ensures that
the G/H current $\widehat{J} = Q_1 \oplus J \oplus Q_3$ is $\mathbb{Z}_4$ graded as $[1,2,3]$ and that infinitely many conserved nonlocal currents can be constructed accordingly.

At quantum level, the supergroup $\widehat{G}$= OSp$(6\vert 2, 2)$ has another nice feature that its Killing form vanishes identically. This means that sigma model on $\widehat{G}$ would be conformally invariant, at least, at one loop. We actually need to quotient $\widehat{G}$ by bosonic subgroup $H$ and consider string worldsheet action on the supercoset $\widehat{G}/H$. This action in general breaks the conformal invariance. To restore the conformal invariance, a suitable Wess-Zumino term needs to be added. It was observed~\cite{Berkovits:1999zq} that the requisite Wess-Zumino term can be constructed provided the bosonic subgroup $H$ is a fixed point set of the $\mathbb{Z}_4$ grading of $\widehat{G}$. This is precisely the same condition that ensures the existence of a flat connection and infinitely many conserved charges thereof. 
Therefore, the supercoset sigma model is conformally invariant and can describe consistent string worldsheet dynamics, at least at one loop order in worldsheet perturbation theory. 

Given such mounting evidences, it is highly likely that Type IIA string on AdS$_4 \times \mathbb{CP}^3$ is integrable at $\lambda \rightarrow \infty$ and further extends to
$\lambda$ finite and even to weak coupling regime~\footnote{We note that the no-go theorem of Goldschmidt and Witten~\cite{Goldschmidt:1980wq} for quantum conservation laws is evaded in the present case since the isotropy subgroup is not simple, and may lead to quantum anomalies~\cite{Abdalla:1982yd}. }.
With such motivation, we now turn to the main part of this work and investigate integrability at the weak coupling regime, $\lambda \rightarrow 0$.
\section{Integrable Spin Chain from Yang-Baxter}
The U(N)$\times \overline{\rm U(N)}$ invariant, single-trace operators under consideration
\bea
&& {{\cal O}^{(I)}}_{(J)} \equiv \mbox{Tr} (Y^{I_1}
Y^\dagger_{J_1} \cdots Y^{I_L} Y^\dagger_{J_L})
\nonumber \\
\simeq
&& {{\cal O}_{(J)}}^{(I)} \equiv \mbox{Tr} (Y^\dagger_{J_1} Y^{I_1}\cdots Y^\dagger_{J_L} Y^{I_L})
\label{spinchainoperator}
\eea
are organized according to SU$_R$(4) irreducible representations. Operator mixing under renormalization and their evolution in perturbation theory can be described by a spin chain of total length $2L$. What kind of spin chain system do we expect? In this section, viewing the operators (\ref{spinchainoperator}) as a spin chain system and utilizing quantum inverse scattering method, we shall derive spin chain Hamiltonian.

As is evident from the structure of operators (\ref{spinchainoperator}), the prospective spin chain involves two types of SU$_R$(4) spins: ${\bf 4}$ at odd
 lattice sites and $\overline{\bf 4}$ at even
 lattice sites. It is thus natural to expect that
the prospective spin chain is
an `alternating SU(4) spin chain' consisting of interlaced
${\bf 4}$ and $\overline{ \bf 4}$. To identify the spin system
and extract its Hamiltonian, it is imperative to solve inhomogeneous Yang-Baxter equations of SU$_R$(4) $\mathfrak{R}$-matrices with varying representations on each site. In fact, a general procedure for solving Yang-Baxter equations in this sort of situations is already set out in \cite{deVega:1991rc}. By construction, resulting spin chain system will be integrable. In this section, we shall follow this procedure and find that the putative SU(4) spin chain is an 'alternating spin chain' involving next-to-nearest neighbor interactions nested with nearest neighbor interactions~\footnote{For a construction in SU(3), see~\cite{abadrios95}. Generalizations
to arbitrary Lie (super)algebras and quantum deformations thereof were studied in~\cite{Aladim:1993uu}-\cite{Ragoucy:2007kg}. }.

We first introduce  ${\mathfrak R}^{\bf 44}(u)$ and
${\mathfrak R}^{\bf 4\bar{\bf 4}}(u)$, where
the upper indices denote SU(4) representations of two spins
involved in `scattering process' and $u, v$ denote spectral parameters. We demand these R-matrices to satisfy two sets of Yang-Baxter equations:
\bea
&& \fR^{\bf 44}_{12}(u-v)\, \fR^{\bf 44}_{13}(u)\, \fR^{\bf 44}_{23}(v)
= \fR^{\bf 44}_{23}(v)\, R^{\bf 44}_{13}(u)\, R^{\bf 44}_{12}(u-v) \label{444YBE} \\
&&
\fR^{\bf 44}_{12}(u-v) \, \fR^{\bf 4\bar{\bf 4}}_{13}(u)\, \fR^{\bf 4\bar{\bf 4}}_{23}(v)
= \fR^{\bf 4\bar{\bf 4}}_{23}(v)\, \fR^{\bf 4\bar{\bf 4}}_{13}(u)\, \fR^{\bf 44}_{12}(u-v)
\label{4bar4bar4YBE} \eea
Here, the lower indices $i,\,j$ denote that the $\fR$ matrix is acting
on $i$-th and $j$-th site $V_i \otimes V_j$ of the full tensor product Hilbert space $V_1 \otimes V_2 \otimes \cdots \otimes V_{2L}$. We easily find that the R-matrices solving (\ref{444YBE}, \ref{4bar4bar4YBE}) are given by
\be
\fR^{\bf 44}(u) = u \mathbb{I} + \mathbb{P} \qquad \mbox{and} \qquad
\fR^{\bf 4\bar{\bf 4}}(u)= -(u+2+\alpha) \mathbb{I} + \mathbb{K}\,
\ee
for an arbitrary constant $\alpha$. Here, we have introduced identity operator $\mathbb{I}$, trace operator $\mathbb{K}$, and permutation operator $\mathbb{P}$:
\be
(\mathbb{I}_{k \ell})^{I_k I_\ell}_{J_k J_\ell} = \delta^{I_k}_{J_k} \delta^{I_\ell}_{J_\ell} \qquad  \qquad
(\mathbb{K}_{k \ell})^{I_k I_\ell}_{J_k J_\ell} = \delta^{I_k I_\ell}
\delta_{J_k J_\ell } \qquad \qquad
(\mathbb{P}_{k \ell})^{I_k I_\ell}_{J_k J_\ell} = \delta^{I_k}_{J_\ell} \delta^{I_\ell}_{J_k}
\ ,
\label{braiding}
\ee
acting as braiding operations mapping tensor product vector space $V_k \otimes V_\ell$ to itself.

We also need to construct another set of R-matrices $\fR^{\bar{\bf 4}\bar{\bf 4}}(u)$ and $\fR^{\bar{\bf 4}\bf 4}(u)$ generating
another alternative spin chain system. We again require them to fulfill the respective
Yang-Baxter equations:
\bea
&& \fR^{\bar{\bf 4}\bar{\bf 4}}_{12}(u-v)\, \fR^{\bar{\bf 4}\bar{\bf 4}}_{13}(u)
\, \fR^{\bar{\bf 4}\bar{\bf 4}}_{23}(v)
= \fR^{\bar{\bf 4}\bar{\bf 4}}_{23}(v)\, \fR^{\bar{\bf 4}\bar{\bf 4}}_{13}(u)\,
\fR^{\bar{\bf 4}\bar{\bf 4}}_{12}(u-v) \label{bar444YBE} \\
&&
\fR^{\bf 44}_{12}(u-v)\, \fR^{\bar{\bf 4}\bf 4}_{13}(u)\, \fR^{\bar{\bf 4}\bf 4}_{23}(v)
= \fR^{\bar{\bf 4}\bf 4}_{23}(v) \, \fR^{\bar{\bf 4}\bf 4}_{13}(u)\,  \fR^{\bf 44}_{12}(u-v)
\label{bar4bar4YBE}
\eea
Again, we find that the R-matrices that solve (\ref{bar444YBE}, \ref{bar4bar4YBE}) are given by
\be
\fR^{\bar{\bf 4}\bar{\bf 4}}(u) = u \mathbb{I} + \mathbb{P} \qquad \mbox{and} \qquad
\fR^{\bar{\bf 4}\bf 4}(u)= -(u+2+ \bar{\alpha})\mathbb{I} + \mathbb{K}\,,
\ee
where 
$\bar{\alpha}$ is an arbitrary constant.

In the two sets of Yang-Baxter equations, the constants $\alpha, \bar{\alpha}$ are undetermined. We shall now
restrict them by requiring unitarity. The unitarity of the combined spin chain system sets the following conditions:
\bea
&& \fR^{\bf 44}(u)\, \fR^{\bf 44}(-u)\  = \rho(u)  \mathbb{I}\, \ \ \nonumber\\
&& \fR^{\bar{\bf 4}\bar{\bf 4}}(u)\, \fR^{\bar{\bf 4}\bar{\bf 4}}(-u)\ = \bar{\rho}(u) \ \mathbb{I} \nonumber\\
&& \fR^{\bf 4\bar{\bf 4}}(u)\, \fR^{\bar{\bf 4}\bf 4}(-u)\  = \sigma(u) \  \mathbb{I}
\eea
where $\rho(u) = \rho(-u), \bar{\rho}(u) = \bar{\rho}(-u), \sigma(u)$ are  $c$-number functions.
It is simple to show that the first two unitarity conditions are indeed satisfied for any $\alpha, \bar{\alpha}$.
It is equally simple to show that the last unitarity condition is is satisfied only if $\alpha= -\bar{\alpha}$. Without loss of generality, in what follows, we shall set $\alpha=- \bar{\alpha}=0$.

Viewing (\ref{spinchainoperator}) as $2L$ sites in a row,
we 
introduce  one transfer T-matrix
%
%
\be
T_0 (u,a)= \fR^{\bf 44}_{01}(u) \fR^{\bf 4{\bar{\bf 4}}}_{02}(u+a)
\fR^{\bf 44}_{03}(u) \fR^{\bf 4{\bar{\bf 4}}}_{04}(u+a)\cdots
\fR^{\bf 44}_{02L-1}(u) \fR^{\bf 4{\bar{\bf 4}}}_{02L}(u+a)\,,
\ee
for one alternate chain and the other T-matrix
\be
\overline{T}_0 (u,\bar{a})
= \fR^{\bar{\bf 4}\bf 4}_{01}(u+\bar{a})
 \fR^{\bar{\bf 4}\bar{\bf 4}}_{02}(u)
 \fR^{\bar{\bf 4}\bf 4}_{03}(u+\bar{a})
\fR^{\bar{\bf 4}{\bar{\bf 4}}}_{03}(u)\cdots
\fR^{\bar{\bf 4}\bf 4}_{02L-1}(u+\bar{a})
\fR^{{\bar{\bf 4}}\bar{\bf 4}}_{02L}(u) \,,
\ee
for the other alternate chain, where
we introduce an  auxiliary zeroth space.
By the standard `train' argument,
one can show that the transfer matrices fulfill the
Yang-Baxter equations,
\be
\fR^{\bf 44}_{00'}(u-v) T_0 (u,a) T_{0'} (v,a)=
 T_{0'} (v,a)  T_{0} (u,a) \fR^{\bf 44}_{00'}(u-v)\,,
\ee
and
\be
\fR^{\bar{\bf 4}\bar{\bf 4}}_{00'}(u-v) \overline{T}_0 (u,\bar{a})
\overline{T}_{0'} (v,\bar{a})=
 \overline{T}_{0'} (v,\bar{a})  \overline{T}_{0} (u,\bar{a})
\fR^{\bar{\bf 4}\bar{\bf 4}}_{00'}(u-v)\,.
\ee
In addition, by a similar argument,
one may verify that
\be
\fR^{{\bf 4}\bar{\bf 4}}_{00'}(u-v+a) {T}_0 (u, {a})
\overline{T}_{0'} (v,-a)=
 \overline{T}_{0'} (v,-a)  {T}_{0} (u,{a})
\fR^{{\bf 4}\bar{\bf 4}}_{00'}(u-v+a)\,.
\ee

We also define the trace of the T matrix by
\be
\tau^{\rm alt}(u,a)=\tr_0 {T}_0 (u, {a})\,.
\ee
and
\be
\overline{\tau}^{\rm alt}(u,\bar{a})
= \tr_0\, \overline{T}_0 (u, \bar{a})
\ee
where the trace is taken over an auxiliary zeroth space.

It then follows from the Yang-Baxter equations that
\bea
&&
[\tau^{\rm alt}(u,a),\tau^{\rm alt}(v,a)]=0
\nonumber\\
&&
[\overline\tau^{\rm alt}(u,\bar{a}),\overline\tau^{\rm alt}(v, \bar{a})]=0\,,
\eea
and
\bea
\hskip1.3cm
[\tau^{\rm alt}(u,a),\bar\tau^{\rm alt}(v, -{a})]=0 \,.
\eea
Here, in the first two equations, $a, \bar{a}$ are arbitrary and
denote two undetermined spectral parameters.
These parameters are restricted further if we demand the last equation
to hold. Indeed, the two alternating transfer matrices
commute each other if and only if  $ \bar{a} = -a$.

As for all other conserved charges, the Hamiltonian is obtained~~\footnote{The following derivation of Hamiltonian is valid only for $L \ge 2$. This means that the energy eigenvalues of the following Hamiltonians
for the case $L=1$ do not agree with true energy eigenvalues.} by evolving the transfer T-matrix infinitesimally in spectral parameter $u$: $H = \rmd \log \tau(u, a)|_{u=0}$ where $\rmd \equiv \partial / \partial u$.
%
%
%
By a straightforward computation,
we obtain the ${\bf 4}\overline{\bf 4}$ spin chain Hamiltonian as
\bea
H_{2\ell - 1}
&=& -(2-a) \mathbb{I} -(4-a^2) \mathbb{P}_{2\ell-1,  2\ell+1} \nonumber \\
&- & (a-2)  \mathbb{P}_{2\ell-1,  2\ell+1}
\mathbb{K}_{2\ell-1,  2\ell}
+(a+2)    \mathbb{P}_{2\ell-1,  2\ell+1}
\mathbb{K}_{2\ell,  2\ell+1} \, ,
\eea
where we scaled the Hamiltonian by multiplying $(a^2-4)$.

By the same procedure, we also find that the Hamiltonian for
for the $\overline{\bf4}{\bf 4}$ spin chain is given by
\bea
\overline{H}_{2 \ell}
&=& -(2+a) \mathbb{I} -(4-a^2) \mathbb{P}_{2\ell,  2\ell+2} \nonumber \\
&+& (a+2)  \mathbb{P}_{2\ell,  2\ell+2}
\mathbb{K}_{2\ell,  2\ell+1}- (a-2)  \mathbb{P}_{2\ell,  2\ell+2}
\mathbb{K}_{2\ell+1,  2\ell+2}
\, ,
\eea
where we have replaced $\bar{a}$ by $a$ using
the relation $\bar{a}=-a$.

At this stage, any choice of the parameter $a$ is possible in
 so far as hermiticity of the Hamiltonian is
satisfied. The latter condition requires that $a$ is a pure
imaginary number. Physically, we are interested in
the situation where ${\bf 4} \leftrightarrow \overline{\bf 4}$
is a symmetry. This is nothing but requiring
charge conjugation symmetry, equivalently,
reflection symmetry in dual lattice. We thus put $a=i 0$~\footnote{Alternatively, one may relax
hermiticity of the Hamiltonian and only demand symmetry under parity and time-reversal, leading to so-called PT-symmetric system~\cite{Bender:1998ke}. This again
sets $a$ to zero. Strictly speaking, however, this latter condition is weaker than the hermiticity
requirement.}
Adding the two alternate Hamiltonians, we get total Hamiltonian~\footnote{We remark the following
useful identities
\bea
\mathbb{P}_{\ell, \ell+2} \mathbb{K}_{\ell, \ell+1} = \mathbb{K}_{\ell+1, \ell+2} \mathbb{P}_{\ell, \ell+2}, \qquad
\mathbb{P}_{\ell,\ell+2} \mathbb{K}_{\ell+1, \ell+2} = \mathbb{K}_{\ell,\ell+1} \mathbb{P}_{\ell, \ell+2} \ .
\label{relations}
\eea
We shall find them useful later when investigating issues concerning wrapping interactions.
}
:
\bea
H_{\rm total} = \sum_{\ell=1}^{2L} H_{\ell, \ell+1, \ell+2}
\eea
with
\bea
H_{\ell, \ell+1, \ell+2}= \Big[ 4\mathbb{I} - 4 \mathbb{P}_{\ell , \ell+2} + 2  \mathbb{P}_{\ell , \ell+2}
\mathbb{K}_{\ell , \ell+1} + 2 \mathbb{P}_{\ell,  \ell+2}
\mathbb{K}_{\ell+1,  \ell+2} \Big] \ .
\label{YBEhamiltonian}
\eea
In this derivation, there is always a freedom of shifting ground state energy by an arbitrary constant. From the outset, we assumed integrability but, except that the symmetry algebra is SU$_R(4)$ and that spins are ${\bf 4}, \overline{\bf 4}$ at alternating lattice sites, we did not utilize any inputs from underlying supersymmetry. With extra input that that supersymmetric ground-state has zero energy, one can always fix the freedom. The (\ref{YBEhamiltonian})
is the Hamiltonian after being shifted by $+6$ per site accordingly.

\section{Integrable Spin Chain from Chern-Simons}
In this section, we approach integrability from weak `t Hooft coupling regime of the superconformal Chern-Simons theory. We use perturbation theory and look for a spin chain Hamiltonian as a quantum part of the dilatation operator acting on the single trace operators.
As mentioned above, in three-dimensional spacetime, general power-counting indicates that logarithmic divergence arises only at even loop orders.
Therefore, leading-order contribution to anomalous dimension starts at two loops.
In general, as well understood from general considerations of the renormalization theory, the divergence in one-particle irreducible diagrams with one insertion of a composite operator contain divergences that are proportional to other composite operators. Therefore, at each order in perturbation theory, all composite operators whose divergences are intertwined must be renormalized simultaneously. In addition, renormalization of elementary fields needs to be taken into account. This leads to the general structure:
\bea
{\cal O}^M_{\rm bare} (Y_{\rm bare}, Y^\dagger_{\rm bare}) = \sum_N {Z^M}_N
{\cal O}^N_{\rm ren} (Z Y_{\rm ren}, Z Y^\dagger_{\rm ren})
\eea
For the operators we are interested in,
this takes the form of
\bea
{\cal O}^M_{\rm bare} = \sum_N {Z^M}_N (\Lambda
) {\cal O}^N_{\rm ren}
\eea
with the UV cut-off scale $\Lambda$.
Therefore,
the anomalous dimension matrix $\Delta$ is given by
\bea
\Delta = {{\rm d} \log Z \over {\rm d} \log \Lambda}.
\eea

In the rest of this section, we compute anomalous dimension matrix for the single trace operators
that were associated with the `alternating spin chain' in the last section:
\bea
{\cal O}^{(I)}_{(J)} = {\rm Tr}\, \Big(
Y^{I_1} {Y}^\dagger_{ J_1 } Y^{I_2} {Y}^\dagger_{J_2}\cdots
Y^{I_L} {Y}^\dagger_{J_L} \Big) \ .
\eea
In ${\cal N}=6$ superconformal Chern-Simons theory, the scalar fields $Y^I,Y^\dagger_I$
are bifundamental fields of U(N)$\times\overline{\rm U(N)}$ gauge group, and transform
as ${\bf 4}$ and $\overline{\bf 4}$ of SU$_R(4)$ R-symmetry group. In Appendix A, we explain
field contents and action of the theory in detail~\footnote{We closely follow notation and convention of~\cite{Benna:2008zy}.}. Schematically, the action of the ABJM theory takes the form
\bea
I = \int_{\mathbb{R}^{2,1}} {k \over 4 \pi} \Big( \mbox{CS}(A)-\mbox{CS}(\overline{A}) \Big)
- \tr (DY)_I^\dagger DY^I + \tr \, \Psi^{I\dagger} i D \hskip-0.23cm / \Psi_I
- V_{\rm F} - V_{\rm B} .
\label{schematicaction}
\eea
Here, the Chern-Simons density is given by
\be
\mbox{CS}(A)= \epsilon^{mnp}\tr \left[
A_m \partial_n A_p + {2i\over 3} A_m A_n A_p \right]\,.
\ee
Covariant derivatives are denoted as $D_m$, while self-interactions involving bosons and fermion pairs
are denoted by $V_{\rm B}, V_{\rm F}$, respectively. See Appendix A for their explicit form. We will recall
them at relevant points in foregoing discussions.

To extract the dilatation operator, we compute the correlation functions
\be
\Big< {\cal O}^{(I)}_{(J)} \mbox{Tr} ({Y}^\dagger_{I_1}Y^{J_1}
\cdots {Y}^\dagger_{I_L} {Y}^{J_L}) \Big> \, \qquad \mbox{for} \qquad L \rightarrow \infty
\ee
by summing over all planar diagrams in perturbation theory in `t Hooft coupling $\lambda$.

In evaluating so, there arises an important issue regarding consistency of regularization with gauge invariance and ${\cal N}=6$ supersymmetry. We shall adopt dimensional reduction method (See, for example, discussions in~\cite{Chen:1992ee}). This method retains $\epsilon^{mnp}$ and Dirac matrices always three-dimensional.  In each Feynman integral, we then manipulate the integrand until all $\epsilon^{mnp}$ and Dirac matrices are eliminated
and the integral is reduced to a Lorentz scalar expression. We then employ dimensional regularization and evaluate the integral. Still, this leaves out infrared divergences that would have been absent were if the theory four-dimensional. As we
will be only concerned with logarithmic ultraviolet divergences, we will take a practical approach that we regularize infrared divergences by introducing mass terms in evaluating Feynman integrals in dimensional regularization. We then remove the regulator mass first and then take the spacetime dimension to three. Previously, it was checked that the dimensional reduction method is consistent with Slavnov-Taylor-Ward identities. Yet, to date, it is not known if the method
is compatible with ${\cal N}=6$ supersymmetry. Thus, in our computations, we shall not assume a priori any input related to supersymmetry. Rather, we will put our result to a test against various consequences of supersymmetry --- for instance, vanishing anomalous dimensions of chiral primary operators and superconformal nonrenormalization theorems.

Using the convention and Feynman rules explained in appendix, we computed all two-loop diagrams that contribute to anomalous dimensions of elementary fields $Y^I, Y^\dagger_I$ and composite operators ${\cal O}^{(I)}_{(J)}$. Acting on the space of the operators, each Feynman diagram can be attributed to the braiding operations $\mathbb{I}$, $\mathbb{K}$, $\mathbb{P}$ introduced in (\ref{braiding}) and their combinations. At two loops, we computed the complete set of Feynman diagrams that contribute to each of these operators. The result turned out
\be
H_{\rm 2-loops}  = {\lambda^2 }
\sum_{\ell = 1}^{2L} \Big[\mathbb{I} - \mathbb{P}_{\ell, \ell+2} +
{1 \over 2} \mathbb{P}_{\ell, \ell+2} \mathbb{K}_{\ell,\ell+1} +
{1 \over 2} \mathbb{P}_{\ell, \ell+2} \mathbb{K}_{\ell+1, \ell+2} \Big]
\label{2loops}
\ee
and this is precisely ${\lambda^2\over 4}$ times the alternating spin chain Hamiltonian (\ref{YBEhamiltonian}) we derived from SU(4) Yang-Baxter equations in the last section.
%
In the rest of this section, we explain essential steps for deriving the Hamiltonian and relegate technical details of evaluating Feynman diagrams in the Appendix. We find it convenient to organize contributing Feynman diagrams according to the number of sites that participate in the Hamiltonian.

\begin{figure}
\begin{center}
\includegraphics[scale=0.7]{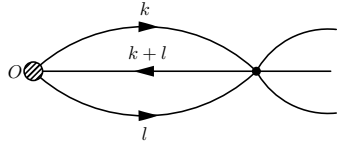}
\end{center}
\caption{\sl Two loop contribution of scalar sextet interaction to anomalous dimension of ${\cal O}$.}
\label{scalarsextet}
\end{figure}

$\bullet$ {\bf Three-site scalar interactions}: \hfill\break
A salient feature of the alternating spin chain Hamiltonian we extracted in section 3 from coupled Yang-Baxter equations is that it contains interactions up to next-nearest-neighbor sites. We thus need to see if such interaction arises from superconformal Chern-Simons planar diagrams and, if so, if the interactions are of the same type.  From the Feynman rules (see Appendix A), it is evident that scalar interaction $-V_{\rm B}$ in (\ref{schematicaction}) is {\sl the} source of three-site interactions, whose explicit form is given by
\bea
V_{\rm B} &=& - {1 \over 3} \left({2 \pi \over k}\right)^2 \overline{\mbox{Tr}} \Big[ \, Y^\dagger_I Y^J Y^\dagger_J Y^K Y^\dagger_K Y^I
+ Y^\dagger_I Y^I Y^\dagger_J Y^J Y^\dagger_K Y^K \nonumber \\
&& \hskip1.8cm + 4 Y^\dagger_I Y^J Y^\dagger_K
Y^I Y^\dagger_J Y^K
- 6 Y^\dagger_I Y^I Y^\dagger_J Y^K Y^\dagger_K Y^J \, \Big]
\label{Bpot}
\eea
The two-loop Feynman diagram is depicted in Fig.\ref{scalarsextet}.
From planar diagram combinatorics of gauge invariant operators at infinite length $2L \rightarrow \infty$, we find the
following contributions arising: $\mathbb{K}_{\ell, \ell+1} +
\mathbb{K}_{\ell+1, \ell+2}$ from the first two terms,
$\mathbb{P}_{\ell, \ell+2}$ from the third  term, and
$\mathbb{I} + \mathbb{P}_{\ell, \ell+2} \mathbb{K}_{\ell, \ell+1}
+ \mathbb{P}_{\ell, \ell+2} \mathbb{K}_{\ell+1, \ell+2}$
from the last term. Taking account of combinatorial
multiplicities, we find that the
scalar sextet potential contributes to the dilatation Hamiltonian as
\be
H_{\rm B} = \lambda^2
\sum_{\ell = 1}^{2L}  \Big[ {1 \over 2} \mathbb{I}-  \mathbb{P}_{\ell,\ell+2} + {1 \over 2} \mathbb{P}_{\ell, \ell+2} \mathbb{K}_{\ell, \ell+1}
+ {1 \over 2} \mathbb{P}_{\ell, \ell+2} \mathbb{K}_{\ell+1, \ell+2}
-  {1 \over 2} \mathbb{K}_{\ell, \ell+1} \ \Big]\,
\ee
(see Appendix B2).
Evidently, compared to the anticipated alternating spin chain Hamiltonian, we have discrepancy in on-site (proportional to $\mathbb{I}$) and nearest neighbor (proportional to $\mathbb{K}$) terms. These are interactions that would arise from gauge or fermion-pair exchange interactions and from wave function renormalization of elementary fields $Y, Y^\dagger$.

\begin{figure}
\begin{center}
\includegraphics[scale=0.8]{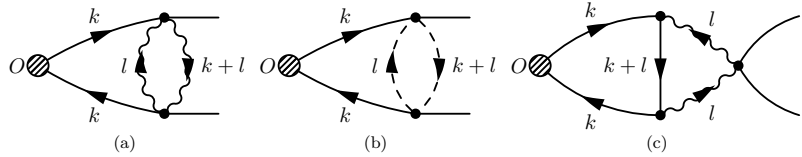}
\end{center}
\caption{\sl Two loop contribution of gauge and fermion exchange interaction to anomalous dimension of ${\cal O}$.}
\label{feyn_05}
\end{figure}

$\bullet$ {\bf Two-site gauge and fermion interactions}: \hfill\break
The scalar fields $Y^I, Y^\dagger_I$ are bifundamentals of U(N)$\times \overline{\rm U(N)}$. Their gauge
interactions can be read off from covariant derivatives:
\bea
D_m Y^I = \partial_m Y^I + i A_m Y^I - i Y^I \overline{A}_m \qquad \mbox{and} \qquad D_m Y^\dagger_I = \partial_m Y^\dagger_I + i \overline{A}_m Y^\dagger_I - i Y^\dagger_I A_m \ .
\eea
As usual, there are paramagnetic interactions (minimal coupling) and diamagnetic interactions (seagull coupling).
We see that gauge interactions contribute to two-site terms for both $\mathbb{I}$ and $\mathbb{K}$. Two
relevant Feynman diagrams are (a) and (c) in Fig.~\ref{feyn_05}.

The Feynman diagram contributing to $\mathbb{I}$ operator arises from square of diamagnetic interactions in $t$-channel. See Fig. \ref{feyn_05}(a). This
diagram is infrared divergent for each subgraphs. We regulate them by giving a mass to internal propagators.
Upon removing the regulator mass to zero, we find a finite part. However, this part turned out ultraviolet convergent and hence does not contribute to anomalous dimension. The
Feynman diagram contributing to $\mathbb{K}$ operator arises from product of diamagnetic interaction and two
paramagnetic interactions. See Fig.~\ref{feyn_05}(c). Taking the net momentum of ${\cal O}$ to zero, which is sufficient for extracting anomalous dimension, we find that only one orientation of diamagnetic interaction vertex yields nonvanishing result.
For details of Feynman rules of gauge interactions and Feynman
diagram evaluation, see Appendix B3.  We found
that gauge interactions contribute to the dilatation operator by
\be
H_{\rm gauge} = {\lambda^2}
\sum_{\ell = 1}^{2L}  \Big[ \ - {1 \over 4}  \mathbb{I} - {1 \over 2}  \mathbb{K}_{\ell, \ell+1} \ \Big].
\ee

Consider next two-site terms induced by fermion-pair exchange diagrams. The relevant part of the Lagrangian in
(\ref{schematicaction}) is the fermion-pair potential:
\bea
V_{\rm F} &=& {2 \pi i \over k} \overline{\mbox{Tr}} \Big[ Y^\dagger_I Y^I \Psi^{\dagger J} \Psi_J
- 2 Y^\dagger_I Y^J  \Psi^{\dagger I} \Psi_J  + \epsilon^{IJKL} Y^\dagger_I \Psi_J Y^\dagger_K \Psi_L \Big]
\nonumber \\
&-& {2 \pi i \over k} \mbox{Tr} \Big[Y^I Y^\dagger_I \Psi_J  \Psi^{\dagger J}
- 2 Y^I Y^\dagger_J \Psi_I \Psi^{\dagger J} + \epsilon_{IJKL} Y^I \Psi^{\dagger J} Y^K \Psi^{\dagger L} \Big] \ .
\label{VF}
\eea
From Feynman rules, we see that planar diagrams formed by square of the second terms in both lines in (\ref{VF}) give rise to $\mathbb{K}$ interactions to the two-loop  dilatation operator. See Fig. \ref{feyn_05}(b) for the relevant Feynman diagram
and Appendix B3 for the details of computation.

In fact, at planar approximation, there is no other
Feynman diagrams that contribute to two-site interactions \footnote{For $L=1$, however, there will be wrapping
interactions. We will discuss them in detail in the next section.}. Taking account of numerical weights in (\ref{VF}),
we find that the fermion potential contributes to the dilatation Hamiltonian as
\be
H_{\rm F}= {\lambda^2} \sum_{\ell = 1}^{2L}  \mathbb{K}_{\ell, \ell+1}
\,.
\ee
%

$\bullet$ {\bf One-site interactions: wave function renormalization} \hfill\break
Adding up all the two-site interactions to the three-site interaction, we see
that terms involving $\mathbb{K}$ operator cancel out one another.
On the other hand,
terms involving $\mathbb{I}$ operator add up to $(1/4) \lambda^2$.
So,
up to overall (volume-dependent) shift of the
ground state energy, the dilatation operator agrees with the alternating spin chain Hamiltonian we derived in the previous section. As we are dealing with superconformal field theory, spectrum of
dilatation generator bears an absolute meaning. Moreover, there could be potential clash between dimensional reduction
we used and superconformal invariance. Therefore, to ensure internal consistency of quantum theory, we shall now compute terms arising from wave function renormalization of $Y, Y^\dagger$. These are all the remaining contributions
to anomalous dimension of composite operator ${\cal O}$.

\begin{figure}
\begin{center}
\includegraphics[scale=0.8]{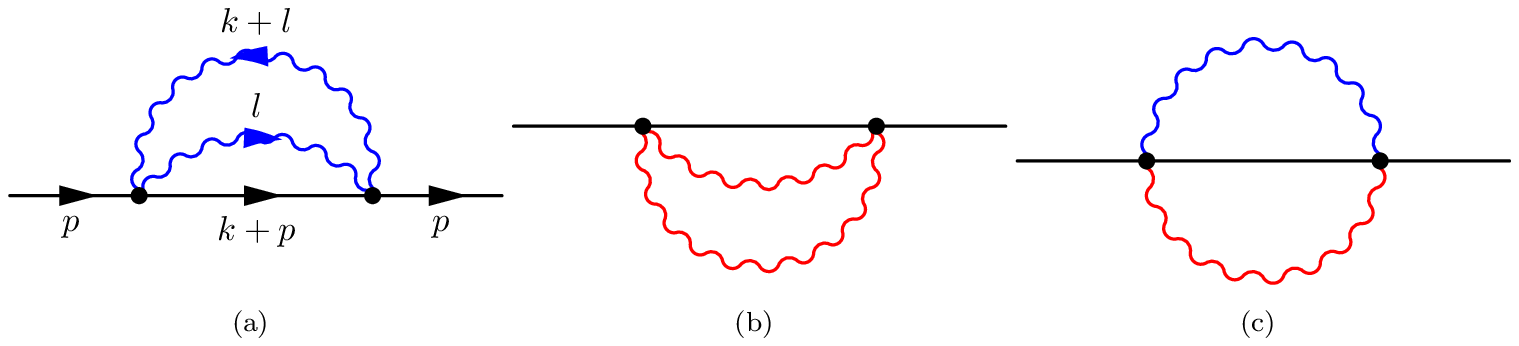}
\end{center}
\caption{\sl Two loop contribution of diamagnetic gauge interactions to wave function renormalization of $Y, Y^\dagger$. They contribute to $\mathbb{I}$ operator in the dilatation operator.}
\label{feyn_01}
\end{figure}

\begin{figure}
\begin{center}
\includegraphics[scale=0.5]{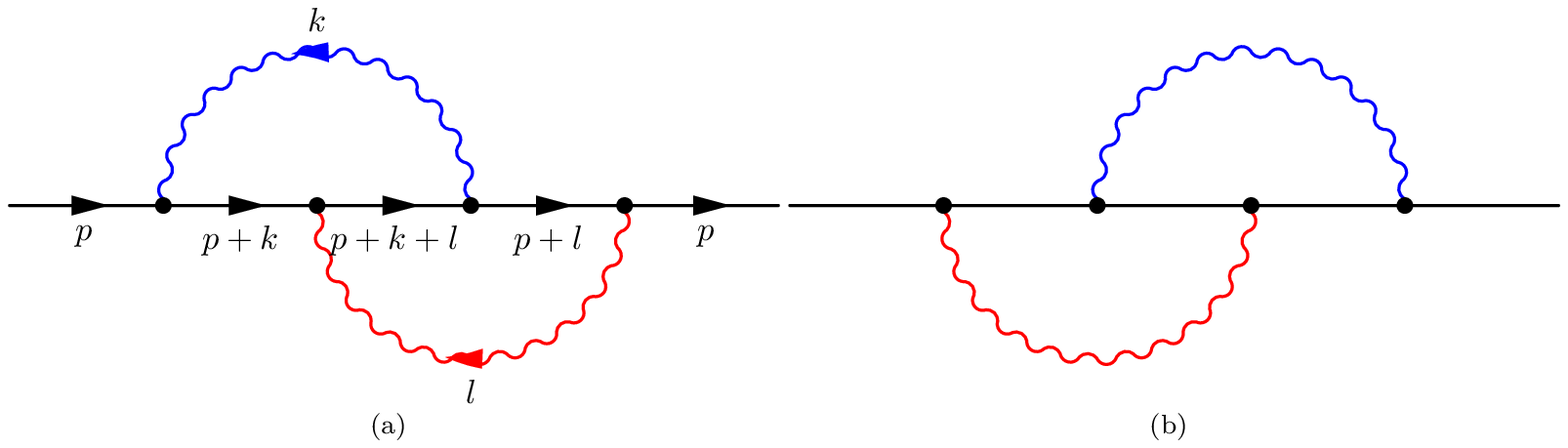}
\end{center}
\caption{\sl Two loop contribution of paramagnetic gauge interactions to wave function renormalization of $Y, Y^\dagger$. They contribute to $\mathbb{I}$ operator in the dilatation operator.}
\label{feyn_02}
\end{figure}

\begin{figure}
\begin{center}
\includegraphics[scale=0.5]{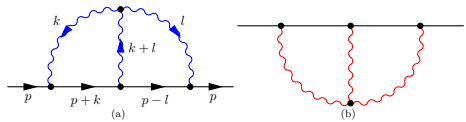}
\end{center}
\caption{\sl Two loop contribution of Chern-Simons interaction to wave function renormalization of $Y, Y^\dagger$.
They contribute to $\mathbb{I}$ operators in the dilatation operator.}
\label{feyn_07}
\end{figure}

Wave function renormalization to $Y, Y^\dagger$ arises
from all three types of interactions. Even though there are
huge numbers of planar Feynman diagrams that could potentially contribute to wave function renormalization,
a vast number of them vanishes identically or cancel one another. First, diagrams involving gauge boson loops
either vanish because of parity-odd nature of the gauge boson propagators or cancel among U(N) and $\overline{\rm U(N)}$ diagrams \footnote{Notice that gauge boson propagator for U(N) and $\overline{\rm U(N)}$ gauge groups
have weight $+k$ and $-k$, respectively.}. Nonzero contribution arise only from diamagnetic interactions shown in
Fig.~\ref{feyn_01},  from paramagnetic interactions shown in Fig.~\ref{feyn_02},
and from Chern-Simons cubic interactions shown in Fig.~\ref{feyn_07}.

\begin{figure}
\begin{center}
\includegraphics[scale=0.8]{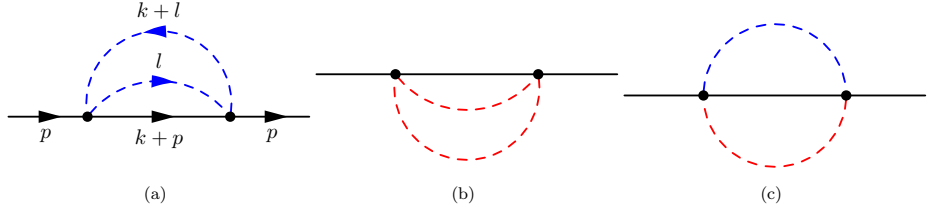}
\end{center}
\caption{\sl Two loop contribution of fermion pair interaction to wave function renormalization of $Y, Y^\dagger$.
They contribute to $\mathbb{I}$ operators in the dilatation operator.}
\label{feyn_03}
\end{figure}

Second, diagrams involving vertices in the first and the second lines in $V_F$ (\ref{VF}) cancel by combinatorics and relative coefficients. Hence, the cancellation is attributable to ${\cal N}=6$ supersymmetry.
The only surviving diagram arise from cross term of vertices in the last line in (\ref{VF}). The Feynman diagram is
shown in Fig.~\ref{feyn_03}.

Third, there are also contributions coming from gauge-matter interactions. Again, almost all diagrams vanish because of parity-odd nature of gauge boson propagator. The only surviving diagrams involve parity-even vacuum polarization, as shown in Fig. \ref{feyn_04}. Their computations are summarized in Appendix B4. We also
present the analysis of the one-loop vacuum polarizations in Appendix
B5.

\begin{figure}
\begin{center}
\includegraphics[scale=0.65]{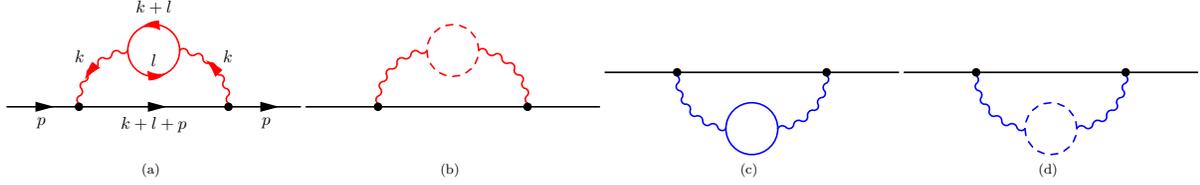}
\end{center}
\caption{\sl Two loop contribution of vacuum polarization to wave function renormalization of $Y, Y^\dagger$.
Both U(N) and $\overline{\rm U(N)}$ gauge parts give additive contributions.}
\label{feyn_04}
\end{figure}

Summing up all these wave function renormalization to  $Y, Y^\dagger$, we find their contribution to the
dilatation operator as
\be
H_{\rm Z} &=& \lambda^2 \Big[ \ \Big({1 \over 12} + {2 \over 3} + {1 \over 3} \Big) + \Big({4 \over 3} + 1 \Big) - {3 \over 8} \Big] \sum_{\ell=1}^{2L} \mathbb{I} \nonumber \\
&=& \lambda^2 \sum_{\ell = 1}^{2L} {3 \over 4} \mathbb{I}
\ee
In the first line, the first parenthesis is the contribution from gauge fields: diamagnetic interactions, paramagnetic interactions, and Chern-Simons interactions. The second parenthesis is the contribution from fermion fields. The last term is the contribution of vacuum polarization. Adding up all the contributions,
\bea
H_{\rm total} &=& H_{\rm B} + H_{\rm F} + H_{\rm gauge} + H_{\rm Z} \eea
we get the result (\ref{2loops}). As claimed, this is precisely the alternating spin chain Hamiltonian we obtained
from mixed set of Yang-Baxter equations. As such, we conclude that dilatation operator of ${\cal N}=6$ superconformal
Chern-Simons theory of ABJM is integrable at two loops.

We stress the importance of explicit and direct computation of the dilatation operator without a prior assumption relying on supersymmetry or integrability. It is satisfying that the result passes various compatibility tests. For instance, take chiral primary operators. These are subset of the single trace operators ${\cal O}$ where $Y$'s and $Y^\dagger$'s are totally symmetric and traceless under any contraction between $Y$'s and $Y^\dagger$'s, and corresponds to massive Kaluza-Klein modes over $\mathbb{CP}^3$ in the Type IIA supergravity dual. Because of supersymmetry, their
scaling dimension should be protected against radiative corrections. Indeed, acting on these operators, $H_{\rm total}$ vanishes since contribution of terms involving $\mathbb{K}$ operator are null and contribution of $\mathbb{P}$ cancel against that of $\mathbb{I}$. As a corollary, the fact that our result is consistent with expectation from supergravity dual implies that the dimension reduction method we adopted for computations are compatible not only with Slavnov-Taylor identities of the gauge symmetry but also with ${\cal N}=6$ supersymmetry.

\section{The Shortest Chain and Wrapping Interactions}
In deriving the dilatation operator in the last section, we assumed that the gauge invariant operator is infinitely long, $L \rightarrow \infty$. From planar diagrammatics, we see easily
that dilatation operator computed perturbatively up to the order $2 \ell$ will give rise to a spin chain Hamiltonian whose range extends to $(2 \ell)$-th order. Therefore, for operators of finite length, a new set of planar diagrams which wraps around the operator will come in to contribute.
These are so-called wrapping interactions, a feature discussed much in the context of integrability of four-dimensional ${\cal N}=4$ super Yang-Mills theory~\cite{Arutyunov:2004vx},~\cite{Sieg:2005kd}-\cite{Fiamberti:2008sh}.

In ${\cal N}=6$ superconformal Chern-Simons theory, the situation is more interesting. Since the
dilatation operator at two loops ranges over three sites, spectrum of the shortest gauge invariant operator of length $2L = 2$ will receive contributions from wrapping diagrams already at leading
order! In this section, we like to identify these wrapping interactions for the shortest gauge invariant operators and discuss their implications.

Let us denote basis of the shortest operators as
\be
|I_1 I_2\rangle =
\tr Y^{I_1} Y^\dagger_{I_2}= O_{I_1 I_2}\, \in {\bf 4} \otimes \overline{\bf 4} \ .
\label{decomposition}
\ee
The ${\bf 4}\otimes {\bf \bar{4}}$ representation is decomposed irreducibly into the traceless
part, $\bf 15$, and the trace part, $\bf{1}$. The multiplet $\bf 15$ is chiral primary operator,
so their conformal dimension ought to be protected by supersymmetry.

To check this, let us first identify the
two-site dilatation operator that includes the wrapping interactions.
At two-loop orders,
the scalar sextet interaction does not contribute to length-$2L=2$ operators since only four legs
can be connected to the operators, leaving a tadpole that vanishes identically.
Hence the dilatation operator consists of the two-site plus wave function
renormalization parts plus wrapping contributions.

From the computations of section 4, the original two-site contributions comprise of
two-loop diagrams from gauge interactions and from $V_{\rm F}$ interactions.
Their contributions are
\be
H_2 =  \Big[\Big(-{1\over 2} \mathbb{K}- {1\over 4} \mathbb{I}\Big)
\lambda^2 + \mathbb{K} \lambda^2 \Big] \times 2
=  \left( \mathbb{K}- {1\over 2} \mathbb{I}\right)  \lambda^2\,.
\ee
In the first line, the first term is the contribution of gauge interaction diagrams
and the second term is the contribution of $V_{\rm F}$ interactions. We computed
total energy, so multiplied the energy density by the spin chain volume $2L=2$.
The one-site contribution arising from the wave function renormalization is
\be
H_1 = {3\over 4} \lambda^2 \,\,\mathbb{I} \times
2
=  {3\over 2} \lambda^2\,\, \mathbb{I} \,.
\label{nonvanishing}
\ee
Adding these two and acting on ${\bf 15}$ in (\ref{decomposition}), we see that anomalous
dimension of the chiral primary operator is non-vanishing. If our regularization method of
dimensional reduction plus infrared mass regularization were compatible with supersymmetry,
there must be other contributions heretofore unaccounted that would cancel against the non-vanishing contribution (\ref{nonvanishing}) and protect the anomalous dimension of chiral primary operator
from quantum corrections. These are precisely wrapping interactions.

Indeed, for the shortest operators of $L=1$ under consideration,
there are three classes of nontrivial wrapping interactions. We now summarize their contribution
and relegate details of Feynman diagram evaluation to Appendix C.

\begin{figure}
\begin{center}
\includegraphics[scale=0.6]{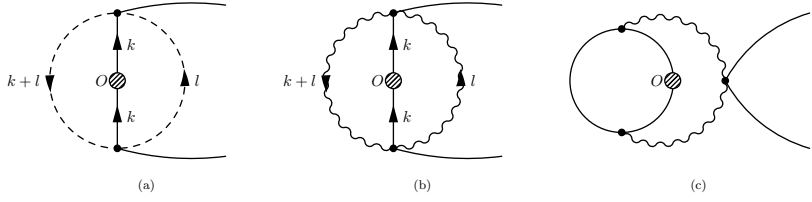}
\end{center}
\caption{\sl Two loop wrapping interaction contribution
to the shortest gauge invariant operators.
(a) fermion field wrapping, (b) gauge field wrapping, (c) a new gauge triangle.}
\label{feyn_08}
\end{figure}
There is the gauge field wrapping contribution with the diamagnetic interactions as
in Fig.~\ref{feyn_08} (a). Its contribution is
\be
H_{gIw}= \lambda^2 \mathbb{I}\,.
\ee
There is also the fermion field wrapping contribution as in
Fig.~\ref{feyn_08} (b). Its contribution is
\be
H_{yw}= 2 \left(\mathbb{K}-\mathbb{I}\right)\lambda^2 \,.
\ee
It is important to note that these two wrapping interactions utilizes simultaneously
U(N) and $\overline{\rm U(N)}$ interactions. Thus, this contribution arises not just
by distinct topology of planar diagram but from very different interactions from the
original, unwrapped two-site interactions.

There is also a doubling-type wrapping contribution
of using the same gauge group interactions. This happens only
for the gauge interaction diagram contributing to $\mathbb{K}$ operator. Moreover,
the contribution is doubled since there are two distinct ways of wrapping. This is
best illustrated on a cylinder, from which we see that there are two different
kinds of topology of wrapped Feynman diagrams.
From appendix C, we identify this contribution as
\be
H_{gKw}= -\lambda^2  \, \mathbb{K}\,.
\ee

Putting both the original and the wrapping diagram
contributions together, the full Hamiltonian of $2L=2$ operator is given by
\be
H_{2L=2} = 2 \,\lambda^2 \,\mathbb{K} \,.
\ee
Notice that the part proportional to $\mathbb{I}$ operator is canceled between
the original and the wrapping interaction contributions.
One thus check that the chiral primary operators $\bf 15$ indeed
has a vanishing anomalous dimension since, by definition, it has no trace part and is annihilated
by $\mathbb{K}$ operator. For the singlet ${\bf 1}$,
$|s\rangle= {1\over 2}|II\rangle$, the anomalous dimension is
\be
H |s\rangle= 8\,\,\lambda^2\,\, |s\rangle\,.
\ee

It is interesting to compare the above spectrum with spectrum of the naive
Hamiltonian $H_{\rm naive}$, viz. the alternating spin chain Hamiltonian
with periodic boundary condition and $2L=2$. The latter is~\footnote{Here, we used the
identities (\ref{relations}).}
\be
H_{\rm naive}=\lambda^2
\sum^2_{\ell=1}\left[\mathbb{I} -  \mathbb{P}_{\ell, \ell+2}
+ {1 \over 2} \mathbb{K}_{\ell+1,\ell+2}  \mathbb{P}_{\ell,\ell+2}
+ {1 \over 2} \mathbb{K}_{\ell,\ell+1}  \mathbb{P}_{\ell, \ell+2}
\right]_{\ell+2 = \ell}
=2 \,\lambda^2\, \mathbb{K}\,,
\ee
for $2L=2$. Acting on ${\bf 15}$ and ${\bf 1}$ states, we find that their anomalous dimension is
$0$ and $4 \cdot 2 \lambda^2$, respectively. So far, we computed the spectrum of the
shortest operators without a priori assumption of supersymmetry. As a consistency check,
we now compare these spectra with their superpartners. Recall that length $2 \ell$ operators with Dynkin labels $(\ell-2m, m+n, \ell-2n)$ and length $2\ell-2$ operators with Dynkin labels
$(\ell-2m, m+n-2, \ell-2n)$ are superpartners each other. Here, we have the simplest situation:
the $L=1$ operator ${\bf 1}$ of Dynkin labels $(0,0,0)$ is the superpartner of $L=2$ operator ${\bf 20}$ of Dynkin labels $(0,2,0)$. Fortuitously, anomalous dimension of the latter was computed at two loops in \cite{Minahan:2008hf} to be
$8 \lambda^2$, and matches perfectly with our computation~\footnote{We thank Joe Minahan and Kostya Zarembo for useful correspondences on this issue.}. Note that, at two loop order, the
$L=2$ operator ${\bf 20}$ does not receive any wrapping interaction corrections. As such, we may
consider agreement of the anomalous dimensions between the two superpartners as a nontrivial
confirmation for the wrapping interactions we studied for the $L=1$ operator ${\bf 1}$. 

We should also note that the naive Hamiltonian is not the right dilatation operator for the shortest operators. Nevertheless, interestingly, the spectrum of naive Hamiltonian coincides with the spectrum extracted from the true two-site Hamiltonian. It would be very interesting to see whether this coincidence persists to higher orders in perturbation theory.

\section{Bethe Ansatz Diagonalization}
In section 3, we constructed transfer matrix. To obtain spectrum, we need to diagonalize the transfer matrices. Within algebraic Bethe ansatz, a fairly general result is known for a Lie (super)groups $G$~\cite{Kulish:1983rd, Aladim:1993uu, Ragoucy:2007kg}. It suffices to adapt the results to the case that $G=$SU(4) \footnote{For SU(3) alternating spin chain, this was done explicitly in \cite{abadrios95}.}. Dynkin diagram of SU(4), drawn horizontally, has three roots: left(l), middle(m), and right(r). The diagonalization is specified by the choice of Dynkin label $(R_l, R_m, R_r)$ for the site representation $R$ and total number of sites $L_R$ that representation occupies. In the present case, we have placed ${\bf 4}$ and $\overline{\bf 4}$ representations at alternating lattices, so $R_l = R_r = 1, R_m = 0$ and $L_{\bf 4} = L_{\overline{\bf 4}} = L$. Each excitation is associated with three sets of Bethe ansatz rapidities $(l_a, m_b, r_c)$'s whose labels range over $[1, N_l], [1, N_m], [1, N_r]$, respectively. It belongs to the SU(4) representation with the Dynkin labels $(L - 2N_l + N_m, N_l + N_r - 2 N_m, L - 2 N_r + N_m)$. Positivity of the Dynkin labels restricts range of the three Bethe ansatz rapidities accordingly. Then, choosing the highest-weight state:
\bea
\vert \Omega_+ \rangle = \prod_{\ell=1}^L \otimes \vert 1 \rangle_{2 \ell - 1} \vert \overline{4} \rangle_{2 \ell} \equiv
\vert 1 \overline{4} 1 \overline{4} \cdots \rangle
\eea
as the ground-state, the eigenvalue of the transfer matrix $T_0(-u)$~\footnote{For later convenience, we choose to diagonalize $T_0$ for opposite sign of the spectral parameter $u$.} is found
to be
\bea
\hskip-0.3cm \Lambda(u) &=& (u-1)^L (u-2)^L \prod_{a=1}^{N_l} {u - i l_a +{1 \over 2} \over u - i l_a - {1 \over 2}} + u^L (u-1)^L \prod_{c=1}^{N_r}{u - i r_c -{5 \over 2} \over u - i r_c-{3 \over 2}} \\
&+& u^L (u-2)^L \Big[ \prod_{a=1}^{N_l} {u - i l_a -{3\over2} \over u - i l_a- {1 \over 2}}
\prod_{b=1}^{N_m} {u - i m_b - 0 \over u - i m_b -1}
+ \prod_{b=1}^{N_m}{u - i m_b -2 \over u - i m_b - 1} \prod_{c=1}^{N_r} {u - i r_c -{1 \over 2} \over u - i r_c -{3 \over 2}}
\Big] \nonumber.
\eea
We have chosen the Bethe rapidities symmetric between the three roots. 
Keeping the highest weight state the same $\vert 1 \overline{4} 1 \overline{4} \cdots \rangle$, we also find that diagonalization of the second transfer matrix $\overline{T}_0(-v)$ proceeds much the same way as that of $T_0(-v)$ except that we
interchange role of the left and the right SU(4) roots:
\bea
\hskip-0.3cm \overline{\Lambda}(v)
&=& v^L (v-1)^L \prod_{a=1}^{N_l} {v - i l_a -{5 \over 2} \over v - i l_a - {3 \over 2}} + (v-1)^L (v-2)^L \prod_{c=1}^{N_r}{v - i r_c +{1 \over 2} \over v - i r_c-{1 \over 2}} \\
&+& v^L (v-2)^L \Big[ \prod_{a=1}^{N_l} {v - i l_a -{1\over2} \over v - i l_a- {3 \over 2}}
\prod_{b=1}^{N_m} {v - i m_b - 2 \over v - i m_b -1}
+ \prod_{b=1}^{N_m}{v - i m_b - 0  \over v - i m_b - 1} \prod_{c=1}^{N_r} {v - i r_c -{3 \over 2} \over v - i r_c -{1 \over 2}}
\Big]. \nonumber
\eea
Mutually commuting conserved charges are then constructed by expanding these eigenvalues around
$u, v=0$. The first two charges are the total momentum and the total energy:
\bea
P_{\rm total} &=& {1 \over i} \Big[\log \Lambda(u) + \log \overline{\Lambda}(u) \Big]_{u=0}
\nonumber \\
&=& \sum_{a=1}^{N_l} \log \left( {l_a + i/2 \over l_a - i/2} \right) + \sum_{b=1}^{N_r} \log
\left( {r_b +i/2 \over r_b - i/2} \right) \ , \label{momentum} \\
E_{\rm total} &=& \lambda^2 \Big[{{\rm d} \over {\rm d} u} (\log \Lambda(u) + \log \overline{\Lambda}(u))\Big]_{u=0}
\nonumber \\
&=& \lambda^2 \Big( \sum_{a=1}^{N_l} { 1 \over l_a^2 + {1 \over 4} } + \sum_{b=1}^{N_r}
{1 \over r_b^2 + {1 \over 4}} \Big) \ .
\label{energy}
\eea
Here, we chose fundamental domain of the momentum to $[ 0, 2 \pi)$ and scaled the total energy
by $\lambda^2$ in accordance to the relation we fixed between Hamiltonian derived from Yang-Baxter
equation and from superconformal Chern-Simons theory. Likewise, we can deduce higher conserved
charges from higher moments of the transfer matrices.

The Bethe equations that results from the above transfer matrix eigenvalues $\Lambda, \overline{\Lambda}$ are
\bea
\left( {l_a - {i \over 2} \over l_a + {i \over 2}} \right)^L &=& \prod_{b=1(b\ne a)}^{N_l}
{l_a - l_b - i \over l_a - l_b + i} \prod_{c=1}^{N_m} {l_a - m_c + {i \over 2} \over l_a - m_c - {i \over 2}} \nonumber \\
1 &=& \prod_{b=1(b\ne a)}^{N_m} {m_a - m_b + i \over m_a - m_b - i}
\prod_{c=1}^{N_l} {m_a - l_c - {i \over 2} \over m_a - l_c + {i \over 2}}
\, \, \prod_{d=1}^{N_r} {m_a - r_d - {i \over 2} \over m_a - r_d + {i \over 2}}
\nonumber \\
\left( {r_a -{i \over 2} \over r_a + {i \over 2}} \right)^L
&=& \prod_{b=1(b\ne a)}^{N_r} {r_a - r_b - i \over r_a - r_b + i}\prod_{c=1}^{N_m} {r_a - m_c + {i \over 2} \over r_a - m_c - {i \over 2}}.
\eea
It is straightforward to check that these same set of Bethe ansatz equations remove potential
simple pole terms for both $\Lambda$ and $\overline{\Lambda}$ simultaneously. 

From the integrability perspectives, $(2+1)$-dimensional superconformal Chern-Simons theory is quite different from $(3+1)$-dimensional super Yang-Mills theory. The most distinct feature is that the spin chain associated with dilatation operator is not homogeneous but alternating. It calls for better understanding to questions that arise in comparison with ${\cal N}=4$ super Yang-Mills
counterpart.  We shall now study spectrum of the Bethe ansatz equations for a few simpler situations and gather features concerning excitations of the alternating spin chain system.

First, consider the special class of $N_m = 0$ for arbitrary $L \ge 2$. The first and the
third Bethe ansatz equations decouple, and each equation becomes the
same as the Bethe ansatz equation of the well-known SU(2) XXX$_{1\over2}$  spin chain.
Thus, if one can identify
the first set with SU(2) of ${\bf 4}$ side, then the third
equation corresponds to SU(2) of $\bar{\bf 4}$.
We then have two decoupled sets of the solution including towers of bound states, and they
are exactly the same as the XXX$_{1\over2}$ spin chain.

Now let us consider the case $N_l=N_m=N_r=1$ case for a general $L \ge 2$. The Bethe ansatz equations are reduced to
\bea
&& \left({l-{i\over 2}\over
l+{i\over 2}}\right)^L ={l-m+{i\over 2}\over
l- m -{i\over 2}}
\nonumber\\
&& \ \ \  m = {1 \over 2} (l + r)
\nonumber\\
&& \left({r-{i\over 2}\over
r+{i\over 2}}\right)^L ={r-m+{i\over 2}\over
r-m -{i\over 2}}\,.
\label{Bethe2}
\eea
In terms of the individual momentum variables, after using the second equation, the combination of
the first and the third equations becomes
\be
e^{i(p_l+p_r)L}=1\, .
\ee
This is solved by
\be
P= p_l + p_r = {2 \pi n\over L},\ \ \ \ \ \ \ \  \ \
(n=0, 1,2, \cdots, L-1)\,.
\ee
First, consider $m=0$ case. In this case, total momentum $P = 0$. For the relative momentum
$q \equiv (p_l - p_r)$,
we also have
\be
e^{i q (L+1)/2} = 1\, .
\ee
This is solved by
\be
{q \over 2} = {2 \pi \mathbb{Z}\over L+1} \, .
\ee
For this case, the energy (\ref{energy}) is given by
\be
E= 8 \lambda^2 \sin^2{q\over 4}.
\ee
This is the simplest example of two-particle excitations where ${\bf 4}$ and $\overline{\bf 4}$
excitations are correlated. The total momentum is zero, while total energy depends on relative
momentum.

Consider next general $m$. From the ratio between the first and the third equations, we obtain
\bea
e^{ i {q \over 2}L} = \mp {l - r - i \over l -r + i} \, ,
\eea
where we used the second Bethe ansatz equation for simplification. Total momentum $P$ is nonzero. Furthermore, expressing this equation
in terms of $P$ and $q$, we find the relations:
\bea
\cos {P \over 2} = {\sin {q \over 4} (L+2) \over \sin {q \over 4} L } \qquad \mbox{or} \qquad
{\cos {q \over 4} (L+2)  \over \cos {q \over 4} L } \, .
\eea
If $q$ is real, viz. two real Bethe roots,
the relation shows that relative momentum $q$ is correlated with total momentum $P$. That is, even though there are two excitations associated with ${\bf 4}$ and $\overline{\bf 4}$ chains, their motion exhibits mutual correlation. If $q$ were imaginary, viz. a Bethe string, the relation shows that total momentum ought to be purely imaginary. This show that there cannot arise any bound-state between ${\bf 4}$ and $\overline{\bf 4}$ spins.

We can also comment on thermodynamic limit in which densities of the Bethe roots are kept finite. By taking $L \rightarrow \infty$ limit of the Bethe ansatz equations and taking the so-called "no hole" excitation condition, we obtain relations among the three Bethe root densities $\rho_l(x), \rho_m (x), \rho_r (x)$. From the first and the third Bethe ansatz equations, after Fourier transform, we find
\bea
\rho_l (k) = \rho_r (k) \, .
\eea
This has a simple interpretation: because the alternating spin chain is manifestly charge-conjugation invariant, excitations ought to be so as well. Moreover, from the second Bethe ansatz equation, we obtain
\bea
\rho_m (k) e^{- {|k|/2}} = {1 \over 2} \Big[ \rho_l (k) + \rho_r (k) \Big].
\eea
It immediately follows from these two equations that the mean value of root densities
\bea
{N_l \over L} = {N_r \over L} \qquad \mbox{and} \qquad { N_m  \over L} = {1 \over 2} \Big(
{N_l \over L} + {N_r \over L} \Big).
\eea
We conclude that all three Bethe root densities are equal, and hence ${\bf 4}$'s and $\overline{\bf 4}$'s are equally populated and balanced each other for the minimum energy configuration.

Furthermore, analysis of the shortest operator suggests that excitation in superconformal Chern-Simons spin chain is different from excitation in ${\cal N}=4$ super Yang-Mills
spin chain. In the latter, the vacuum is ferromagnetic and excitations break SO$_R$(6)
to [SU(2)]$^2$. The latter is the symmetry group of dilute, finite-energy excitations.
In the present case, analysis of the previous section seems to indicate that excitation is
organized by the full SU(4), not by any subgroup of it. This is because the finite energy excitation is a singlet of SU(4), not of any subgroup of it. Lastly, in this system, excitations with $N_m=0$ comprises of two decoupled XXX$_{1\over2}$ spin chains with its own ferromagnetic vacuum, respectively. Though this is certainly a closed subsector, general excitations in the full system looks quite different, as is seen above in the simple situation of $N_m=1$. 

Following the general prescription~\cite{Ragoucy:2007kg} and paving the parallels to what was done in the context of ${\cal N}=4$ super Yang-Mills theory \cite{Beisert:2003yb}, extending the SU(4) spin chain to the OSp$(6 \vert 2, 2; \mathbb{R})$ superspin chain and writing down Bethe ansatz equations are immediate and straightforward. This was done already in \cite{Minahan:2008hf}. More recently, spectrum in the Penrose limit~\cite{Nishioka:2008gz}, various SU($2\vert 2$) closed subsectors~\cite{Gaiotto:2008cg}, all loop Bethe ansatz equations~\cite{Gromov:2008qe}, and finite-size effects~\cite{Grignani:2008te} were studied. With these developments, it would be interesting to explore precision tests for the new correspondence proposed by ABJM.

\section*{Acknowledgement}
We would like to thank Dongmin Gang for extensive Mathematica check on an issue related to integrability and to David Berenstein, Hyunsoo Min,
Joe Minahan, Takao Suyama, Satoshi Yamaguchi, Kostya Zarembo for correspondences and discussions.
This work was supported in part by R01-2008-000-10656-0 (DSB),
SRC-CQUeST-R11-2005-021 (DSB,SJR), KRF-2005-084-C00003 (SJR), EU FP6 Marie Curie Research \& Training Networks MRTN-CT-2004-512194 and HPRN-CT-2006-035863 through MOST/KICOS (SJR), and F.W. Bessel Award of Alexander von Humboldt Foundation (SJR).
S.J.R. thanks the Galileo Galilei Institute for Theoretical Physics for hospitality during the course of this work.

\appendix

\section{Notation, Convention and Feynman Rules}
\subsection{Notation and Convention}
$\bullet$ $\mathbb{R}^{1,2}$ metric:
\bea
&& g_{mn} = \mbox{diag}(-, +, +) \quad \mbox{with} \quad  m,n = 0, 1, 2. \nonumber \\
&& \epsilon^{012} = - \epsilon_{012} = +1 \nonumber \\
&& \epsilon^{mpq} \epsilon_{mrs} = - (\delta^p_r \delta^q_s - \delta^p_s \delta^q_r); \qquad
\epsilon^{mpq} \epsilon_{mpr} = - 2 \delta^q_r \nonumber \\
\eea
$\bullet$ $\mathbb{R}^{1,2}$ Majorana spinor and Dirac matrices:
\bea
&& \psi \equiv \mbox{two-component} \,\,\, \mbox{Majorana} \,\,\, \mbox{spinor} \nonumber \\
&& \psi^\alpha = \epsilon^{\alpha \beta}\psi_\beta ,
\quad \psi_\alpha = \epsilon_{\alpha \beta} \psi^\beta \quad
\mbox{where} \quad \epsilon^{\alpha \beta} = - \epsilon_{\alpha \beta} = i \sigma^2 \nonumber \\
&& {\gamma^m_\alpha}^\beta = ( i \sigma^2, \sigma^3, \sigma^1),
\quad (\gamma^m)_{\alpha \beta} = (-\mathbb{I}, \sigma^1, -
\sigma^3) \quad \mbox{obeying} \quad \gamma^m \gamma^n = g^{mn} -
\epsilon^{mnp} \gamma_p. \eea

\subsection{ABJM Theory}
$\bullet$ Gauge and global symmetries:
\bea
&& \mbox{gauge symmetry}: \quad \mbox{U(N)} \otimes \overline{\mbox{U(N)}} \nonumber \\
&& \mbox{global symmetry}: \quad \mbox{SU(4)}
\eea
We denote trace over U(N) and $\overline{\rm U(N)}$ as Tr and $\overline{\rm Tr}$, respectively.

$\bullet$ On-shell fields are gauge fields, complexified Hermitian scalars and Majorana spinors ($I=1,2,3,4$):
\bea
&& A_m: \quad \mbox{Adj}\,\,\, (\mbox{U(N)}); \hskip2cm \overline{A}_m : \quad
\mbox{Adj}\,\,\, \overline{\mbox{U(N)}} \nonumber \\
&& Y^I = (X^1 + i X^5, X^2 + i X^6, X^3 - i X^7, X^4 - i X^8):  \qquad ({\bf N}, \overline{\bf N}; {\bf 4}) \nonumber \\
&& Y^\dagger_I = (X^1 - i X^5, X^2 - i X^6, X^3 + i X^7, X^4 + i X^8): \hskip0.8cm ( \overline{\bf N}, {\bf N}; \overline{\bf 4})
\nonumber \\
&& \Psi_I = (\psi^2 + i \chi^2, - \psi^1 - i \chi^1, \psi_4  - i \chi_4 , - \psi_3 + i \chi_3 ) : \hskip0.4cm
({\bf N}, \overline{\bf N}; \overline{\bf 4}) \nonumber \\
&& \Psi^{\dagger I} = (\psi_2 - i \chi_2, - \psi_1 + i \chi^1, \psi^4 + i \chi^4, - \psi^3 - i \chi^3): \hskip0.3cm
(\overline{\bf N}, {\bf N}; {\bf 4})
\eea
$\bullet$ action:
\bea
I &=&   \int_{\mathbb{R}^{1,2}}
\Big[ \, {k \over 4\pi} \epsilon^{mnp} \mbox{Tr} \left(A_m \partial_n A_p +{2 i \over 3} A_m A_n A_p \right)
- {k \over 4\pi} \epsilon^{mnp} \overline{\mbox{Tr}} \left(  \overline{A}_m \partial_n \overline{A}_p +{2 i \over 3} \overline{A}_m
\overline{A}_n \overline{A}_p \right) \nonumber \\
&& \hskip1cm +{1 \over 2} \overline{\mbox{Tr}} \left( -(D_m Y)^\dagger_I D^m Y^I  + i \Psi^{\dagger I} D \hskip-0.22cm / \Psi_I  \right) + {1 \over 2} \mbox{Tr} \left(- D_m Y^I (D^m Y)^\dagger_I  +
 i \Psi_I D \hskip-0.22cm / \Psi^{\dagger I}  \right)
\nonumber \\
&&
\hskip1cm - V_{\rm F} - V_{\rm B} \, \Big]
\eea
Here, covariant derivatives
are defined as
\bea
D_m Y^I = \partial_m Y^I + i A_m Y^I - i Y^I \overline{A}_m \, , \quad D_m Y^\dagger_I = \partial_m Y^\dagger_I + i \overline{A}_m Y^\dagger_I - i Y^\dagger_I A_m
\eea
and similarly for fermions $\Psi_I, \Psi^{\dagger I}$. Potential terms are
\bea
V_{\rm F} &=& {2 \pi i \over k} \overline{\mbox{Tr}} \Big[ Y^\dagger_I Y^I \Psi^{\dagger J} \Psi_J
- 2 Y^\dagger_I Y^J  \Psi^{\dagger I} \Psi_J  + \epsilon^{IJKL} Y^\dagger_I \Psi_J Y^\dagger_K \Psi_L]
\nonumber \\
&-& {2 \pi i \over k} \mbox{Tr} [Y^I Y^\dagger_I \Psi_J  \Psi^{\dagger J}
- 2 Y^I Y^\dagger_J \Psi_I \Psi^{\dagger J} + \epsilon_{IJKL} Y^I \Psi^{\dagger J} Y^K \Psi^{\dagger L} \Big]
\eea
and
\bea
V_{\rm B} &=& - {1 \over 3} \left({2 \pi \over k}\right)^2
\overline{\mbox{Tr}} \Big[ \, Y^\dagger_I Y^J Y^\dagger_J Y^K Y^\dagger_K Y^I
+ Y^\dagger_I Y^I Y^\dagger_J Y^J Y^\dagger_K Y^K \nonumber \\
&& \hskip1.8cm + 4 Y^\dagger_I Y^J Y^\dagger_K
Y^I Y^\dagger_J Y^K -
6 Y^\dagger_I Y^I Y^\dagger_J Y^K Y^\dagger_K Y^J \, \Big]
\eea
At quantum level, since the Chern-Simons term shifts by integer multiple of
$8 \pi^2$, not only $N$ but also $k$
should be integrally quantized.
To suppress the cluttering $2\pi$ factors, we also use the notation
$\kappa= {k\over 2\pi}$.
At large $N$, we expand the theory and
physical observables in double series of
\bea
g_{\rm st} = {1 \over N}, \qquad \lambda = {N \over k} =
 {N \over 2 \pi\kappa}
\eea
by treating them as continuous perturbation parameters.

\subsection{Feynman Rules}
$\bullet$ We adopt Lorentzian Feynman
rules and manipulate all Dirac matrices and $\epsilon_{mnp}$
tensor expressions to scalar integrals.
For actual evaluation of these integrals,
we shall go the Euclidean space integral by the Wick rotation,
which corresponds to $x^0 \rightarrow -i\tau$. In the momentum
space, this means we change the contour of $p_0$
 to the imaginary axis 
following
the standard Wick rotation. Then in terms
of integration measure, we simply replace
${\rm d}^{2 \omega} k \rightarrow
 i {\rm d}^{2 \omega} k_{\rm E}$ together
 with $p^2 \rightarrow  + p_{\rm E}^2$.
The procedure is known to obey Slavnov-Taylor
identity, at least to two loop order.
\hfill\break
$\bullet$ We choose covariant gauge fixing condition for both gauge groups:
\bea
\partial^m A_m = 0 \qquad \mbox{and} \qquad \partial^m \overline{A}_m = 0
\eea
and work in Feynman gauge by setting the gauge parameter $\xi$ to unity. Accordingly , we introduce a pair of
Faddeev-Popov ghosts $c, \overline{c}$ and their conjugates, and add to $I$ the ghosts action:
\bea
I_{\rm ghost} = \int_{\mathbb{R}^{2,1}} \Big[ \mbox{Tr} \partial^m c^* D_m c + \overline{\mbox{Tr}}
\partial^m  \overline{c}^* D_m \overline{c} \Big]
\eea
Here, $D_m c = \partial_m c + i [A_m, c]$ and $D_m \overline{c} = \partial_m \overline{c} + i
[\overline{A}_m, \overline{c}]$.

$\bullet$ Propagators in U(N)$\times\overline{\mbox{U(N)}}$ matrix notation:
\bea
\mbox{gauge propagator}: \quad && \Delta_{mn}(p) ={2 \pi \over k} \mathbb{I} \,  {\epsilon_{mnr} p^r \over p^2 - i \epsilon} \nonumber \\
\mbox{scalar propagator}: \quad && {D_I}^J (p) = \delta_I^J \,  {-i \over p^2 - i \epsilon}  \nonumber \\
\mbox{fermion propagator}: \quad && {S^I}_J (p) =\delta^I_J \, { i p \hskip-0.22cm / \over p^2 - i \epsilon}   \nonumber \\
\mbox{ghost propagator}: \quad && K(p) \, = \, {-i \over p^2 - i \epsilon}
\eea
$\bullet$ Interaction vertices are obtained by multiplying $i = \sqrt{-1}$ to nonlinear terms of the Lagrangian density. Note that the paramagnetic coupling of gauge fields to scalar fields has the invariance property under simultaneous exchange between $A_m, Y^I$ and $\overline{A}_m, Y^\dagger_I$.

\section{Two-Loop Computations}

\subsection{Two-loop integrals}
We first tabulate various Feynman
integrals that appear recurrently among two-loop diagrams.
They are all evaluated straightforwardly by Feynman parametrization
\bea
{1 \over A^a B^b} = {\Gamma(a+b) \over \Gamma(a) \Gamma(b)}
\int\int {\rm d} x {\rm d} y \delta(1 - x - y) {x^{a-1} y^{b-1} \over (Ax + By)^{a+b}}.
\eea
We use dimensional regularization by shifting the spacetime dimension to $d = 2 \omega = 3 - \epsilon$. The ultraviolet divergence shows up as a simple pole $1/ \epsilon$. It is related to the momentum space cutoff $\Lambda$ as
\bea
{1 \over \epsilon} := 2 \log \Lambda\,.
\eea
In the following, we collect factors arising from propagators
in parenthesis and those from vertices  in square bracket.
We have the following integrals:
\bea
\hskip-3cm \bullet \qquad \quad I_1 &=& \int {{\rm d}^{2\omega} k \over (2 \pi)^{2\omega}} {{\rm d}^{2\omega} \ell \over (2 \pi)^{2\omega}} {1 \over (k+\ell)^2} {1 \over k^2} {1 \over \ell^2}
\nonumber \\
&=& \int_0^1 {\rm d} x \int {{\rm d}^{2 \omega} \ell \over (2 \pi)^{2\omega}} {1 \over \ell^2}
\int {{\rm d}^{2 \omega} k \over (2 \pi)^{2 \omega}}
{1 \over [k^2 + 2 x k \cdot \ell + x \ell^2 ]^2} \nonumber \\
&=& -{1 \over 8 \pi} \int_0^1 {\rm d} x {1 \over \sqrt{x (1 - x)}}
\int {{\rm d}^{2 \omega} \ell \over (2 \pi)^{2 \omega}} {1 \over \sqrt{ (k^2)^3}} \nonumber \\
&=&  + {1 \over 8} {1 \over 4 \pi^2} {1 \over \epsilon}.
\eea

The integral that appears in fermion and gauge boson exchange diagrams is:
\bea
\hskip-4cm \bullet \qquad \quad I_2 &=&
\int {{\rm d}^{2\omega} k \over (2 \pi)^{2\omega}}
{{\rm d}^{2\omega} \ell \over (2 \pi)^{2\omega}}
{1 \over (k^2)^2} {2 (k+\ell)\cdot \ell
 \over (k+\ell)^2 \,\, \ell^2}\,.
\eea
We perform the $\ell$ integral first after using the Feynman
reparametrization:
\bea
\hskip-3cm  \qquad \quad I_2
&=& \int_0^1 {\rm d} x
\int {{\rm d}^{2 \omega} k \over (2 \pi)^{2 \omega}}
{1 \over (k^2)^2}
\int {{\rm d}^{2 \omega}
\ell \over (2 \pi)^{2\omega}}
{2 \ell \cdot (\ell +k) \over [\ell^2 + 2 x k \cdot \ell + x k^2 ]^2} \nonumber \\
&=& - {1 \over 8 \pi} \int_0^1 {\rm d} x { \sqrt{x} \over \sqrt{ 1 - x}}
\int {{\rm d}^{2 \omega} k \over (2 \pi)^{2 \omega}}
{1 \over k^3} \nonumber \\
&=&  - {1 \over 8} {1 \over 4 \pi^2} {1 \over \epsilon}\,,
\label{Itwo}
\eea
where for the second equality, we have used the integral,
\bea
&&  \int {{\rm d}^{2 \omega} \ell \over (2 \pi)^{2 \omega}}
{\ell_m \ell_n \over [\ell^2 + 2x \ell \cdot k + k^2]^2}
=
{1 \over (4 \pi)^{3/2}} \Big[
 { x^2 k_mk_n \Gamma(1/2) \over [x(1-x) k^2]^{1/2}} + {g_{mn} \over 2} {\Gamma(-1/2)
\over [x(1-x) k^2]^{-1/2}}\Bigr]\nonumber\\
&&
\int {{\rm d}^{2 \omega} \ell \over (2 \pi)^{2 \omega}}
{\ell_m  \over [\ell^2 + 2x \ell \cdot k + k^2]^2}
= -{1 \over (4 \pi)^{3/2}}
 { x k_m \Gamma(1/2) \over [x(1-x) k^2]^{1/2}} \,.
\eea

If one exchanges the order of integrations, there may appear an
infrared singularity.
However by introducing
infrared regulator mass $m$, one may get the same result
in the limit $\omega \rightarrow 3/2$ and $m \rightarrow 0$.

In the gauge boson exchange diagram, the following integral appears:
\bea
\hskip1.5cm \bullet \qquad \quad I_3 = \int {{\rm d}^{2 \omega} k \over (2 \pi)^{2 \omega}} \int {{\rm d}^{2 \omega} \ell
\over (2 \pi)^{2 \omega}} {1 \over (k+\ell)^2} {1 \over (k^2)^2} {1 \over (\ell^2)^2}
[ (k \cdot \ell)^2 - k^2 \ell^2] \nonumber \equiv I_{3,A} - I_{3,B}
\eea
We evaluated them as follows:
\bea
I_{3,A} &=& \int {{\rm d}^{2 \omega} k \over (2 \pi)^{2 \omega}} {k_m k_n \over (k^2)^2}
\int_0^1 {\rm d} x {\Gamma(3) \over \Gamma(2)} \int {{\rm d}^{2 \omega} \ell \over (2 \pi)^{2 \omega}} {(1 - x) \ell_m \ell_n \over [\ell^2 + 2x \ell \cdot k + k^2]^3} \nonumber \\
&=& \int {{\rm d}^{2 \omega} k \over (2 \pi)^{2 \omega}} {k_m k_n \over (k^2)^2}
\int_0^1 {\rm d} x (1 - x) {1 \over (4 \pi)^{3/2}} \Big[ x^2 k_mk_n {\Gamma(3/2) \over [x(1-x) k^2]^{3/2}} + {g_{mn} \over 2} {\Gamma(1/2) \over [x(1-x) k^2]^{1/2}} \Big] \nonumber \\
&=& {1 \over 16} {1 \over 4 \pi^2} {1 \over \epsilon}. \nonumber \\
I_{3,B} &=& \int {{\rm d}^{2 \omega} \over (2 \pi)^{2 \omega}} {1 \over k^2} \int_0^1 {\rm d} x
\int {{\rm d}^{2 \omega} \ell \over (2 \pi)^{2 \omega}}
{1 \over [\ell^2 + 2 x \ell \cdot k + x k^2]^2} \nonumber \\
&=& {\Gamma(1/2) \over (4 \pi)^{3/2} \Gamma(2)} \int_0^1 {\rm d} x {1 \over x (1 - x)}
\int {{\rm d}^{2 \omega} k \over (2 \pi)^{2 \omega}} {1 \over (k^2)^{3/2}} \nonumber \\
&=& {1 \over 8} {1 \over 4 \pi^2} {1 \over \epsilon}
\eea
Hence,
\bea
I_3 = \Big( {1 \over 16} - {1 \over 8} \Big) {1 \over 4 \pi^2} {1 \over \epsilon} = - {1 \over 16} {1 \over 4 \pi^2} {1 \over \epsilon}.
\eea
%

\subsection{Contribution from Sextet Scalar Potential}
The Lagrangian contains sextet scalar interaction $-V_{\rm scalar}$. Three of the scalar fields couple to ${\cal O}$ and the rest three to
${\cal O}^\dagger$. With U(N) and
$\overline{\rm U(N)}$ index loops, combinatorial
factors are given by
\bea
-3 \cdot N^2  \Big[ 2 \mathbb{I}^{\otimes^3} - 4 \mathbb{P}_{13} \otimes \mathbb{I}_2 - \mathbb{K}_{12} \otimes \mathbb{I}_3 - \mathbb{I} \otimes \mathbb{K}_{23}  \otimes \mathbb{I}_2 + 2 \mathbb{K}_{13} \otimes \mathbb{K}_{12} + 2 \mathbb{K}_{12} \otimes \mathbb{K}_{13} \Big]
\eea
There are three scalar propagators and one
interaction vertex, contributing factors
\bea
{1 \over 3 \kappa^2} (i)^3\,\,[i\,] \cdot N^2 = {4 \pi^2 \over 3} \lambda^2
\eea
The remaining 2-loop integral is given by $I_1$. Summing
over all contributions, the scalar
sextet interaction gives rise to 2-loop dilatation operator
\bea
H_{\rm B} = {\lambda^2 \over 2} \sum_{\ell=1}^{2L} \Big[ \mathbb{I} - 2 \mathbb{P}_{\ell, \ell+2} - \mathbb{K}_{\ell, \ell+1} + \mathbb{P}_{\ell, \ell+2} \mathbb{K}_{\ell, \ell+1} + \mathbb{K}_{\ell, \ell+1}
 \mathbb{P}_{\ell, \ell+1} \Big]
\eea

\subsection{Contribution from two-site Interactions}

In this appendix we shall present the full detailed
computation of the two site interactions.
First let us compute the Yukawa two-site interactions. The nonvanishing
Yukawa interaction leads to only a $\mathbb{K}$-type
interaction. The relevant Feynman diagram is depicted in
Fig.~\ref{feyn_05}b.

With two Yukawa interaction components
and one U(N) and one
$\overline{\rm U(N)}$ color traces, combinatorial factors are gathered as
\bea
{1 \over 2!} \cdot 2 \cdot N^2 = N^2.
\eea
There are four propagators and two vertices. This yields numerical factors
\bea
(-i)^2 (i)^2 [i\,]^2 \left( \pm {2i \over \kappa 
} \right)^2 \cdot (-)_{\rm FD}
\eea
where the subscript $( \,\,)_{\rm FD}$ signifies the Fermi-Dirac
statistics minus sign. 
The loop integral is given by
\bea
  i^2 \int {{\rm d}^{2 \omega} k \over (2 \pi)^{2 \omega}}
{{\rm d}^{2 \omega} \ell \over (2 \pi)^{2 \omega}}
{1 \over (-(k)^2)^2}\mbox{tr}
\Big({ \ell\hskip-0.23cm / \over \ell^2} {k \hskip-0.23cm /
+\ell \hskip-0.23cm /
  \over (k+\ell)^2} \Big)
\label{yukawaK1}
\eea
where the $i^2$ factor comes from the analytic continuation of the
integration measure.

After taking the gamma matrix trace tr$\gamma^m \gamma^n = 2 g^{mn}$,
this integral equals to $-I_2$ in (\ref{Itwo}).

Hence putting everything together, one has
\be
\lambda^2 {(-1)\over 2\epsilon} \mathbb{K}
\ee
for the Yukawa two-site interactions.
The contribution to the operator renormalization
is negative of this: Therefore, the Yukawa contribution
is
\bea
{H}_{\rm F} = \lambda^2 \sum_{\ell=1}^{2L}
\mathbb{K}_{\ell, \ell+1}
\eea

We now evaluate the gauge two-site interactions.

The gauge boson interactions contribute both
$\mathbb{K}$ and $\mathbb{I}$ type diagrams to the dilatation operator.
Let us begin with $\mathbb{K}$ type contribution. The
relevant diagram is
in Fig.~\ref{feyn_05}c. It has combinatorial factors
\bea
{1 \over 2!} \cdot 2 \cdot N^2 = N^2.
\eea
There are three boson propagators, two gauge propagators and one seagull
interaction vertex. So, numerical factors are given by
\bea
(-i)^3 \cdot [-i\,]^3 \cdot (\pm 1)^2 \Big( {1 \over \kappa 
} \Big)^2 = - {4 \pi^2 \over k^2}
\eea
where the last factor accounts for the $(\pm)$ relative sign of
U(N) and $\overline{\rm U(N)}$
Chern-Simons term. It is important to note that the gauge field propagator
in momentum space has no $i = \sqrt{-1}$. The loop integral reads
\bea
i^2
\int {{\rm d}^{2 \omega} k \over (2 \pi)^{2 \omega}}
{{\rm d}^{2 \omega} \ell \over (2 \pi)^{2 \omega}}
{1 \over (k + \ell)^2} {1 \over (k^2)^2} {1 \over (\ell^2)^2}
(\epsilon_{mnp} (k + 2 \ell)^n k^p) g^{mq} (\epsilon_{qrs} (k + 2 \ell)^r (- k)^s)
\eea
where again the $i^2$ factor comes from the Euclidean rotation.
Using the identity $g^{mq} \epsilon_{mnp} \epsilon_{qrs}
= - (g_{nr} g_{ps} - g_{ns} g_{pr})$, we
find that the integral is the same as $4 I_3$.

Hence putting everything together, one has
\be
-{\lambda^2\over 2} {(-1)\over 2\epsilon} \mathbb{K}
\ee
for the gauge two site $\mathbb{K}$ contributions and,
for the operator renormalization,
\be
-{\lambda^2\over 2} {1\over 2\epsilon} \mathbb{K}\,.
\ee

There are also contributions to $\mathbb{I}$ from $t$-channel
exchange of diamagnetic gauge boson
interaction. The corresponding Feynman diagram is depicted in
Fig~\ref{feyn_05}a.
There are two scalar propagators, two gauge boson propagators
and two diamagnetic vertices. Note again,
for  Chern-Simons theory,
gauge boson propagator
has no $i$ in momentum space. So, the combinatorial factor is
\bea
{1 \over 2!} 2 \cdot (-i)^2 \cdot [i\,]^2
\cdot N^2 \cdot \Big({1 \over \kappa 
}\Big)^2  = (4 \pi^2) \lambda^2.
\eea
The loop integral reads
\bea
i^2
\int {{\rm d}^{2 \omega} k \over (2 \pi)^{2 \omega}}
{{\rm d}^{2 \omega} \ell \over (2 \pi)^{2 \omega}}
{1 \over (k^2)^2} {\epsilon^{mna} (k+\ell)_a \over (k+\ell)^2}
{{\epsilon_{mn}}^b \ell_b \over \ell^2}
\label{gaugeI1}
\eea
Using the identity $\epsilon^{mna} {\epsilon_{mn}}^b = -2 g^{ab}$,
we find that this integral is
the same as $-I_2$. There are identical contributions
from each letter (with alternating U(N) and
$\overline{\rm U(N)}$ gauge boson exchanges), we find the contribution
 as
\bea
-{\lambda^2\over 4} {(-1)\over 2\epsilon} \mathbb{I}\,.
\eea
 The corresponding operator renormalozation contribution is
\bea
-{\lambda^2\over 4} {1\over 2\epsilon} \mathbb{I}\,.
\eea

Hence there are two gauge  two-site contributions. Using
 $1/(2\epsilon) = \ln \Lambda$, the gauge two-site contributions to
the anomalous dimension  are summarized as
\bea
H_{\rm gauge}=\sum^{2L}_{\ell =1} \left[-{1\over 4} \mathbb{I} - {1\over 2}
\mathbb{K}_{\ell, \ell+1} \right] \lambda^2\,.
\eea

\subsection{ Contributions of  Wave Function Renormalization}

The first one involves diamagnetic gauge interactions. The relevant
Feynman diagrams are in Fig.~\ref{feyn_01}.
As scalar fields
are bi-fundamentals, there are processes involving U(N) gauge boson
pair, $\overline{\rm U(N)}$ gauge boson pair, and one U(N) gauge boson
and one $\overline{U(N)}$ gauge boson pair, which are respectively
corresponding to Fig.~\ref{feyn_01}a, Fig.~\ref{feyn_01}b and
Fig.~\ref{feyn_01}c.
Taking account of opposite
relative sign between gauge boson propagators for U(N) and
$\overline{\rm
U(N)}$
and of different combinatorial weight of diamagnetic coupling terms,
the numerical factor reads
\bea
{1 \over 2!} 2 \cdot (-i) \cdot [i\,]^2
 \cdot [ (-)^2 \cdot (+)^2 + (+)^2 \cdot (-)^2
+ (+)(-) \cdot (-2)^2] N^2 \Big({1 \over \kappa 
} \Big)^2 = - 2i (4 \pi^2) \lambda^2
\eea
Denote momentum of the external scalar field as $p^m$. Then, loop integral reads
\bea
i^2
\int {{\rm d}^{2 \omega} k \over (2 \pi)^{2 \omega}}
{{\rm d}^{2 \omega} \ell \over (2 \pi)^{2 \omega}}
{1 \over (k+\ell+p)^2} { \epsilon^{mna} k_a \over k^2}
{{\epsilon_{mn}}^b \ell_b \over \ell^2}\,.
\eea
For the evaluation of this integral, let us introduce
\bea
&& I_G (p)=
\int {{\rm d}^{2 \omega} k \over (2 \pi)^{2 \omega}}
{{\rm d}^{2 \omega} \ell \over (2 \pi)^{2 \omega}}
{1 \over (k+\ell+p)^2} { 2 k\cdot \ell \over  k^2\,\, \ell^2}
\nonumber \\
&=&  2 \int {{\rm d}^{2 \omega}
\ell \over (2 \pi)^{2 \omega}} {1 \over \ell^2} \int_0^1
{\rm d} x \int {{\rm d}^{2 \omega} k \over (2 \pi)^{2 \omega}}
{ k \cdot \ell \over [(k + x (\ell+ p))^2 + x(1-x)(\ell+p)^2]^2} \nonumber \\
&=& -{1 \over 4\pi} \int_0^1 {\rm d} x \sqrt{x \over 1 - x} \int
{{\rm d}^{2 \omega}
\ell \over (2 \pi)^{2 \omega}}
{\ell \cdot (\ell + p) \over \sqrt{(\ell + p)^2}}.
\eea

The $x$-integral is finite and equals to $\pi/2$. The remaining
$\ell$-integral can be performed by applying Feynman's parametrization.
In dimensional regularization, we have
\bea
&& -{1 \over 8} {\Gamma(3/2) \over \Gamma(1/2)}
\int_0^1 {\rm d} y {1 \over \sqrt{y}}
\int {{\rm d}^{2 \omega} \ell \over (2 \pi)^{2 \omega}}
{ \ell \cdot p \over [\ell^2 + 2 x \ell \cdot p + x p^2]^{3/2}}
 \nonumber \\
&&= -{1 \over 16} \int_0^1 {{\rm d} y \over \sqrt{y}} \Big[ -
{\Gamma(\epsilon) \over (4 \pi)^\omega \Gamma(3/2)}
{y p^2 \over (y(1-y) p^2)^\epsilon} \Big]
\eea
Taking $\epsilon = 3/2 - \omega \rightarrow 0$, this integral equals to
\bea
I_G = {1 \over 24} {1 \over 4 \pi^2} {1 \over \epsilon}.
\eea
Putting together, we thus find that these diagrams contribute
to the wave function renormalization as
\bea
 -{1 \over 12} \lambda^2 {1 \over \epsilon} (ip^2)
\eea

Consider next two diagrams involving four paramagnetic couplings.
Planar diagrams involve two
vertices from U(N) and two from $\overline{\rm U(N)}$,
as shown in Fig. \ref{feyn_02}. Taking
care of opposite relative sign of gauge boson propagators
between U(N) and $\overline{U(N)}$
and that there are three internal scalar propagators,
we have combinatorial factors
\bea
{1 \over (2!)^2} 2^2 \cdot (+)(-)
\cdot (-i)^3 [i\,]^2 \Big({1 \over \kappa 
} \Big)^2 N^2  (2) = - 2 i (4 \pi^2) \lambda^2.
\eea
where we put an additional factor two because there are two such diagrams.
With external momentum $p^m$, the loop integral read
\be
i^2
\int {{\rm d}^{2 \omega} k \over (2 \pi)^{2 \omega}}
{{\rm d}^{2 \omega} \ell \over (2 \pi)^{2 \omega}}
{
 \epsilon^{mnq} (\ell + 2p)_m (p+ 2k +2\ell)_n \ell_q
\epsilon^{abc} (2\ell + k+2p)_a (k+2p)_b k_c
 \over (k+\ell+p)^2  (\ell+p)^2  (k+p)^2  k^2 \ell^2} \,.
\ee
This integral can be integrated without further
assumption but we note that the numerator of the integrand is already
quadratic in $p_m$. Using the isotropy of the system, we replace
\be
p_a p_b \  \ \rightarrow \ \ {p^2\over 3} g_{ab}
\ee
and then set $p$ to zero in the remaining integral.
One may show that the results
from the both methods agree precisely with each other.

Thus the integral becomes
\be
- {16\over 3} p^2 \, \,\, I_3 = {p^2\over 12 \pi^2} {1\over \epsilon}\,.
\ee
Putting all the factor together, one has
\bea
 -{2 \over 3} \lambda^2 {1 \over \epsilon} (ip^2)
\eea

There are two diagrams involving Chern-Simons cubic coupling.
The contributions of U(N) and $\overline{\rm U(N)}$
are added up with an equal weight. The Feynman diagrams are
in Fig.~\ref{feyn_07}.

The relevant combinatorics is
\bea
{3!\cdot 3\over 3!}(-i)^2 [i\,]^4 N^2
\Big({(\pm)\kappa i \over 3 
} \Big)
\Big({(\pm 1) \over \kappa 
} \Big)^3  (2) = -
  2i (4 \pi^2) \lambda^2
\eea
where the last factor two takes care of
the $\overline{\rm U(N)}$ contribution.
The loop integral becomes
\be
i^2
\int {{\rm d}^{2 \omega} k \over (2 \pi)^{2 \omega}}
{{\rm d}^{2 \omega} \ell \over (2 \pi)^{2 \omega}}
{
 \epsilon^{mqn} \epsilon_{mar} (2p+k)^a k^r
\epsilon_{nbs} (2p-\ell+k)^b (-k-\ell)^s
\epsilon_{qct} (2p-\ell)^b \ell^t
 \over (p-\ell)^2  (k+\ell)^2  (k+p)^2  k^2 \ell^2} \,.
\ee
Using the rule of
$$p_ap_b \rightarrow {\delta_{ab}\over 3} p^2\,,$$
the integral becomes
\be
i^2 {8p^2\over 3}
\int {{\rm d}^{2 \omega} k \over (2 \pi)^{2 \omega}}
{{\rm d}^{2 \omega} \ell \over (2 \pi)^{2 \omega}}
{-k^2 \ell^2 +(k\cdot \ell)^2
 \over (k+\ell)^2  (k^2)^2  (\ell^2)^2} \,.
\ee
Using $I_3$, one finds
\be
{p^2\over 6} {1\over 4\pi^2} {1\over \epsilon} \,.
\ee
Therefore, the whole contribution combining the combinatorics
becomes
\be
-{1\over 3} {\lambda^2}  {1\over \epsilon} {ip^2}\,.
\ee

There are also
diagrams involving paramagnetic and diamagnetic couplings.
Their net combinatorial factor is nonzero, but the
loop integral vanishes identically.

Let us now turn to the Yukawa contributions. First consider the
Feynman diagrams in Fig.~\ref{feyn_03}a and Fig.~\ref{feyn_03}b.
Within the planar diagram, both
fermion can be joined either U(N) side (Fig.~\ref{feyn_03}a)
and $\overline{\rm U(N)}$ side (Fig.~\ref{feyn_03}b).
The joining using the first two terms has a factor $4$ from the SU(4)
index contraction. Then the cross terms between the first two and
the second two terms  in total have a factor $-4$. Hence one can check that
this cross contributions cancel precisely the those from the first two.

By combining the second two of the Yukawa potential,
for the U(N) and $\overline{\rm U(N)}$
side, we have combinational factors
\bea
{1 \over 2!} 2 \cdot (-i\cdot i^2) [i\,]^2
\cdot  \Big({2i \over \kappa 
} \Big)^2 N^2 (-)_{\rm FD}
\times 8  = - 32 i (4 \pi^2) \lambda^2.
\eea
where the extra factor eight comes from one
contraction of SU(4) index and
the doubling by  U(N) and $\overline{\rm U(N)}$.

Then the remaining integral has the expression,
\bea
  i^2 \int {{\rm d}^{2 \omega} k \over (2 \pi)^{2 \omega}}
{{\rm d}^{2 \omega} \ell \over (2 \pi)^{2 \omega}}
{1 \over (p+k-\ell)^2}{\mbox{tr} \, \,  \ell \hskip-0.23cm / \ \,
k \hskip-0.23cm /
  \over k^2\,\,  \ell^2}
\eea
which is same as $I_G$. Therefore the whole contribution is
\bea
 -{4 \over 3} \lambda^2 {1 \over \epsilon} (ip^2)\,.
\eea

For the wave function renormalization, the third two of Yukawa potential
also contribute.
The diagram is in Fig.~\ref{feyn_03}c.
It has combinatoric factors,
\bea
 (2!)^2  \cdot (-i)\cdot (i)^2 [i\,]^2
\cdot  \Big({i \over \kappa 
} \Big) \Big({-i \over \kappa 
} \Big) N^2 (-)_{\rm FD}
\times (-6)  = - 24 i (4 \pi^2) \lambda^2\,,
\eea
where $(2!)^2$ is the usual symmetry factor of the Feynman diagram.  The
last factor $(-6)$ comes from the following $SU(4)$ index contraction
\be
\epsilon_{IABC}\epsilon^{JCBA} = -6 \,\, \delta_I^J
\ee
where $I$ is for the incoming and the $J$ for the outgoing scalar
$SU(4)$ indices.

Then the remaining integral takes precisely
the same from:
\bea
  i^2 \int {{\rm d}^{2 \omega} k \over (2 \pi)^{2 \omega}}
{{\rm d}^{2 \omega} \ell \over (2 \pi)^{2 \omega}}
{1 \over
(p+k-\ell)^2}{\mbox{tr} \, \,  \ell \hskip-0.23cm / \ \, k \hskip-0.23cm /
  \over k^2\,\,  \ell^2}
\eea
which is again the same as $I_G$. Therefore the whole contribution is
\bea
 - \lambda^2 {1 \over \epsilon} (ip^2)\,.
\eea

Finally,
there are the vacuum polarization contributions of the gauge loop.
The relevant diagrans are depicted in Fig.~\ref{feyn_04}.

As we shall explain in the following appendix,
the self energy correction for both $A$ and $\overline{A}$ gauge
fields is given by
\be
i\, \Pi_{ab}(k)= 8  i
\Bigl[ { { k_a  k_b- g_{ab} k^2  }\over 16 k}\Bigr]\,,
\ee
where the factor eight comes from the four complex scalars and fermions
with an equal weight.

For the relevant diagram of Fig.~\ref{feyn_04}, the combinatorics factor
reads
\bea
 {2!\over 2!}  \cdot (-i) [i\,]^2
\cdot  \Big({1 \over \kappa 
} \Big)^2  N^2
\times (2)  =  2 i (4 \pi^2) \lambda^2\,,
\eea
where the last factor two comes from the doubling by
replacing $A$ gauge by the $\overline{A}$ gauge field.
The remaining Feynman integrals  takes the from,
\bea
&& i \int {{\rm d}^{2 \omega} k \over (2 \pi)^{2 \omega}}
{
 \epsilon^{amn} (2p+ k )_m (-k)_n
\epsilon^{bij} (k+2p)_i k_j \, i \,\Pi_{ab} (k)
 \over (k+p)^2  (k^2)^2} \,,\nonumber\\
&& = 2 \int {{\rm d}^{2 \omega} k \over (2 \pi)^{2 \omega}}
{ k^2 p^2 - (k\cdot p)^2
 \over (k+p)^2  \, \, k^3}
\eea
where we have a single $i$ produced by the Euclidean rotation.

By the dimensional regularization,
this leads to
\be
   p^2  {1\over 3\pi^2} {1\over \epsilon}\,.
\ee

Hence the total contribution reads
\be
 {8\over 3} \lambda^2 {1 \over \epsilon} (ip^2)\,.
\ee

All the remaining diagrams, one may prove
that their contribution is identically
zero after the dimensional regularization.

Finally we add up all the above contributions to the
wave function renormalization and  find that
\be
 -{3\over 4} \lambda^2 {1 \over \epsilon} (-)(-ip^2)\,.
\ee

Since the counter term is a negative of this, the two-loop
scalar wave function renormalization becomes
\be
Z_s = 1 -{3\over 4} \lambda^2 {1 \over \epsilon}
= 1  -{3\over 4}  \lambda^2 (2\ln \Lambda) \,.
\label{wave}
\ee
In order to get the operator renormalization factor,
one has to take $Z_s^{1\over 2}$ out for each site,
which corresponds to adding $-{1\over 2}$ of (\ref{wave}) to
the interaction part of renormalization. The final contribution
to the anomalous dimension is
\be
H_Z = \lambda^2
\sum^{2L}_{\ell=1}\left[\left({1\over 12}+{2\over 3}+{1\over 3}\right)
+\left({4\over 3}+1\right)   -{8\over 3}\right]
 \mathbb{I}= \lambda^2\sum^{2L}_{\ell=1}
{3\over 4}  \mathbb{I}\,.
\ee

For the gauge two-loop contributions $1/12$, $2/3$, $1/3$
including the gauge self-energy correction contribution $8/3$,
the two-loop Feynman diagram computation is carried out
in Ref.\cite{dasilva} for the $U(1)$ case. One can check the
precise agreement after taking care of the planarity factor
and the number of matter degrees. Furthermore, Ref.~\cite{dasilva2}
deals with the two-loop Yukawa contribution to the scalar wave
function renormalization for again $U(1)$. This result
is again matching with ours if one takes care of
the planarity
and the number of fermions.

\subsection{One loop self energy correction to
the gauge field}

The self-energy correction enters in the same form for
the U(N) and the $\overline{\rm U(N)}$ gauge fields.
Therefore we focus on the correction to $A$ gauge
field only. At the one-loop level,
the boson, the fermion, the gauge and the ghost loops
may in general contribute to the gauge self-energy correction.
In this appendix, we identify these self-energy contributions.

We begin with the scalar loop contribution. It is the sub-diagram of
Fig.~\ref{feyn_04}a. The momentum $k$ plays the role of the external momentum.
The self energy contribution reads
\bea
 i\, \Pi^s_{ab} (k)= (i)^2 [i\,]^2 (4)i
 \int {{\rm d}^{2 \omega} \ell \over (2 \pi)^{2 \omega}}
{ (2\ell +k)_a (2\ell+k)_b
  \over (k+\ell)^2\,\,  \ell^2}\,,
\eea
where the extra factor $4$ comes from the fact that
4 complex scalars are coupled to the gauge field.
Using the dimensional regularization, one obtains
\be
i\,\Pi^s_{ab} (k)= (4)i\Bigl[  {{ k_a  k_b -g_{ab} k^2 }\over 16 k}\Bigr]\,.
\ee

Similarly, for the fermion loop, the self-energy
contribution becomes
\bea
 i\, \Pi^f_{ab} (k)= (i)^2 [i\,]^2 (4)(-)_{\rm FD}
 \, i \int {{\rm d}^{2 \omega} \ell \over (2 \pi)^{2 \omega}}
{ \tr \gamma_a \, (\ell\hskip-0.23cm / +
k\hskip-0.23cm / \, ) \,\gamma_b\,  \ell \hskip-0.23cm /
  \over (k+\ell)^2\,\,  \ell^2}\,,
\eea
where again the extra factor four comes from the fact that there are
4 complex fundamental fermions. Using the $\gamma$ matrix identity and the
 dimensional regularization, the contribution becomes
 \be
i\, \Pi^f_{ab} (k)= (4) i\Bigl[  {{ k_a  k_b -g_{ab} k^2 }\over 16k}\Bigr]\,.
\ee
Hence, each complex matter contributes by the same weight and sign.

One can continue the dimensions $2\omega$ to four and obtain the vacuum polarization
in four-dimensional Yang-Mills theories. The integration leads to the logarithmic
divergence in this case contributing positively to the $\beta$-function of the Yang-Mills
coupling. Again, boson and fermion contributions add up.

For the gluon self-energy contribution, we have
 \bea
i\, \Pi^A_{ab} (k) &=& (3)\cdot (3) [i^2]
\Bigl[{i\kappa \over 3}\Bigr]^2 \Bigl[{1\over \kappa}\Bigr]^2
(i)^2 [i\,]^2 (4) i
 \int {{\rm d}^{2 \omega} \ell \over (2 \pi)^{2 \omega}}
{ \epsilon^{mbn} \epsilon^{jai} \epsilon_{imq} \epsilon_{njr} (\ell+k)^q \ell^r
  \over (k+\ell)^2\,\,  \ell^2}\,,\nonumber\\
&=&
 i \int {{\rm d}^{2 \omega} \ell \over (2 \pi)^{2 \omega}}
{(\ell +k)_a \ell_b + (\ell +k)_b \ell_a
  \over (k+\ell)^2\,\,  \ell^2}\,.
\eea
It becomes
\be
i\, \Pi^A_{ab} (k)=  -i\Bigl[  {{ k_a  k_b +g_{ab} k^2 }\over 32 k}\Bigr]\,,
\ee
which alone does not respect the gauge invariance.
However, there exists also the ghost loop contribution,
\bea
i\, \Pi^{\rm gh}_{ab} (k)
=(i)^2 [i\,]^2 (-)i \int {{\rm d}^{2 \omega} \ell \over (2 \pi)^{2 \omega}}
{(\ell +k)_a \ell_b + (\ell +k)_b \ell_a
  \over (k+\ell)^2\,\,  \ell^2}\,,
\eea
where we put the extra ($-$) sign due to the ghost statistics.
Therefore, the ghost contribution cancels out precisely
the gauge loop contribution, reproducing the well-established result~\cite{Chen:1992ee}.

Again, analytically continuing to four dimensions, the integral expression for the
gauge part changes while the ghost integral remains intact.
With Yang-Mills couplings, both contributions no longer cancel each other but
contribute negatively to the $\beta$-function.

\section{Wrapping Interactions for the Two-Sites}

As in Fig.~\ref{feyn_08}, there are occuring three kinds of
wrapping interactions.
First is the gauge
interactions of two diamagnetic couplings in Fig.~\ref{feyn_08}b.
It is an $\mathbb{I}$ type interaction and happens, not for each
site, but just once.

The combinatorial factor is
\bea
{1 \over 2!} 2 \cdot (-i)^2 \cdot [2 i\,]^2 \cdot (+)(-)
\cdot N^2 \cdot \Big({1 \over \kappa
}\Big)^2  = - (4) {4 \pi^2 \lambda^2}.
\eea
where (+)(-) accounts for the the relative
$U(N)$ and $\overline{U(N)}$
Chern-Simons term and the factor two in the vertices
takes care of the diamagnetic interaction.

The loop integral
\bea
i^2
\int {{\rm d}^{2 \omega} k \over (2 \pi)^{2 \omega}}
{{\rm d}^{2 \omega} \ell \over (2 \pi)^{2 \omega}}
{1 \over (k^2)^2} {\epsilon^{mna} (k+\ell)_a \over (k+\ell)^2}
{{\epsilon_{mn}}^b \ell_b \over \ell^2}
\eea
is the same as (\ref{gaugeI1}).
So, the loop integral is evaluated as
\be
{1\over 8} {1\over 4\pi^2} {1\over\epsilon }\,.
\ee
Putting things together, we find the whole contribution
as
\bea
{\lambda^2} {(-1)\over 2\epsilon} \mathbb{I}\,.
\eea
 The corresponding operator renormalization contribution is
\bea
{\lambda^2} {1\over 2\epsilon} \mathbb{I}\,.
\eea

The second is for the $K$ type gauge wrapping,
whose Feynman diagram is depicted in Fig.~\ref{feyn_08}c.
It is doubling
of the K-type interaction discussed
for the general two-site gauge interactions. This
doubling occurs due to the fact that, on the
cylinder, one may have two different topology
of the diagrams. Namely the diamagnetic
interaction of the same gauge group
may happen either one side
or the other side, which is not possible for the
infinite chains.
From the previous result, the corresponding extra
operator renormalization contribution is
\bea
-{\lambda^2} {1\over 2\epsilon} \mathbb{K}\,,
\eea
where we take into account of the fact that
the doubling occurs both for the $U(N)$ and
$\overline{U(N)}$.

There is an additional wrapping interaction coming from
the third two terms in the Yukawa potential. The Feynman diagram
is in Fig.~\ref{feyn_08}.
In order to have proper contractions, one has to join operator site
one
($Y^{I_1}$) to the site two ($Y^{J_2}$) whereas the operator site
two $Y^\dagger_{I_2}$ to $Y^\dagger_{J_1}$. The corresponding
$\epsilon$ tensors in the Yukawa interaction produce
\be
\epsilon_{I_1 A J_2 B}\, \epsilon^{I_2 B J_1 A}
= 2 (\mathbb{I}- \mathbb{K})\,.
\ee

The  combinatorial factors are gathered as
\bea
{2! \over 2!} \cdot 2 \cdot 2
(-i)^2 (i)^2 [i\,]^2
\cdot N^2 \cdot
\left( {i \over \kappa 
} \right)
\left( {-i \over \kappa 
} \right)
\cdot (-)_{\rm FD}
= 4  (4\pi^2)\cdot \lambda^2
\,.
\eea

The loop integral is given by
\bea
  i^2 \int {{\rm d}^{2 \omega} k \over (2 \pi)^{2 \omega}}
{{\rm d}^{2 \omega} \ell \over (2 \pi)^{2 \omega}}
{1 \over (-(k)^2)^2}\mbox{tr}
\Big({ \ell\hskip-0.23cm / \over \ell^2} {k \hskip-0.23cm /
+\ell \hskip-0.23cm /
  \over (k+\ell)^2} \Big)\,,
\eea
which is the same as (\ref{yukawaK1}).
By the loop integration, one gets
\be
 {1 \over 8} {1 \over 4 \pi^2} {1 \over \epsilon}\,.
\ee

Hence putting everything together, one has
\be
2 \lambda^2 {(-1)\over 2\epsilon} (\mathbb{K}-\mathbb{I})
\ee
for the Yukawa wrapping interactions.
Therefore, the Yukawa contribution to the operator
renormalization is
\be
2 \,\lambda^2 {1\over 2\epsilon} (\mathbb{K}-\mathbb{I})\,.
\ee

Adding up the gauge and Yukawa contributions,
the wrapping interaction contribution to the two-site Hamiltonian
is
\be
H_{\rm wrap}=\mathbb{I}- \mathbb{K} +2(\mathbb{K}-\mathbb{I}) =
-\mathbb{I} + \mathbb{K}\,.
\ee


\begin{thebibliography}{99}

\bibitem{Aharony:2008ug}
  O.~Aharony, O.~Bergman, D.~L.~Jafferis and J.~Maldacena,
  ``N=6 superconformal Chern-Simons-matter theories, M2-branes and their
  gravity duals,''
  arXiv:0806.1218 [hep-th].

\bibitem{Maldacena:1997re}
  J.~M.~Maldacena,
  Adv.\ Theor.\ Math.\ Phys.\  {\bf 2} (1998) 231
  [Int.\ J.\ Theor.\ Phys.\  {\bf 38} (1999) 1113]
  [arXiv:hep-th/9711200].

\bibitem{Nilsson:1984bj}
  B.~E.~W.~Nilsson and C.~N.~Pope,
  Class.\ Quant.\ Grav.\  {\bf 1} (1984) 499.

\bibitem{Minahan:2002ve}
  J.~A.~Minahan and K.~Zarembo,
  JHEP {\bf 0303} (2003) 013
  [arXiv:hep-th/0212208].

\bibitem{Beisert:2003tq}
  N.~Beisert, C.~Kristjansen and M.~Staudacher,
  Nucl.\ Phys.\  B {\bf 664} (2003) 131
  [arXiv:hep-th/0303060].

\bibitem{Frolov:2003qc}
  S.~Frolov and A.~A.~Tseytlin,
  Nucl.\ Phys.\  B {\bf 668} (2003) 77
  [arXiv:hep-th/0304255].

\bibitem{Beisert:2003jj}
  N.~Beisert,
  Nucl.\ Phys.\  B {\bf 676} (2004) 3
  [arXiv:hep-th/0307015].

\bibitem{Beisert:2003yb}
  N.~Beisert and M.~Staudacher,
  Nucl.\ Phys.\  B {\bf 670} (2003) 439
  [arXiv:hep-th/0307042].

\bibitem{Arutyunov:2003uj}
  G.~Arutyunov, S.~Frolov, J.~Russo and A.~A.~Tseytlin,
  Nucl.\ Phys.\  B {\bf 671} (2003) 3
  [arXiv:hep-th/0307191].

\bibitem{Beisert:2003ys}
  N.~Beisert,
  Nucl.\ Phys.\  B {\bf 682} (2004) 487
  [arXiv:hep-th/0310252].

\bibitem{Beisert:2004hm}
  N.~Beisert, V.~Dippel and M.~Staudacher,
  JHEP {\bf 0407} (2004) 075
  [arXiv:hep-th/0405001].

\bibitem{Arutyunov:2004vx}
  G.~Arutyunov, S.~Frolov and M.~Staudacher,
  JHEP {\bf 0410} (2004) 016
  [arXiv:hep-th/0406256].

\bibitem{Beisert:2004ry}
  N.~Beisert,
  Phys.\ Rept.\  {\bf 405} (2005) 1
  [arXiv:hep-th/0407277].

\bibitem{Staudacher:2004tk}
  M.~Staudacher,
  JHEP {\bf 0505} (2005) 054
  [arXiv:hep-th/0412188].

\bibitem{Beisert:2005bm}
  N.~Beisert, V.~A.~Kazakov, K.~Sakai and K.~Zarembo,
  Commun.\ Math.\ Phys.\  {\bf 263} (2006) 659
  [arXiv:hep-th/0502226].

\bibitem{Beisert:2005fw}
  N.~Beisert and M.~Staudacher,
  Nucl.\ Phys.\  B {\bf 727} (2005) 1
  [arXiv:hep-th/0504190].

\bibitem{Beisert:2005tm}
  N.~Beisert,
  ``The $su(2|2)$ dynamic S-matrix,''
  arXiv:hep-th/0511082.

\bibitem{Hofman:2006xt}
  D.~M.~Hofman and J.~M.~Maldacena,
  J.\ Phys.\ A  {\bf 39} (2006) 13095
  [arXiv:hep-th/0604135].

\bibitem{Beisert:2006ib}
  N.~Beisert, R.~Hernandez and E.~Lopez,
  JHEP {\bf 0611} (2006) 070
  [arXiv:hep-th/0609044].

\bibitem{Beisert:2006qh}
  N.~Beisert,
  J.\ Stat.\ Mech.\  {\bf 0701} (2007) P017
  [arXiv:nlin/0610017].

\bibitem{Beisert:2006ez}
  N.~Beisert, B.~Eden and M.~Staudacher,
  J.\ Stat.\ Mech.\  {\bf 0701} (2007) P021
  [arXiv:hep-th/0610251].

\bibitem{Gaiotto:2007qi}
  D.~Gaiotto and X.~Yin,
  JHEP {\bf 0708} (2007) 056
  [arXiv:0704.3740 [hep-th]].

\bibitem{Brezin:1979am}
  E.~Brezin, C.~Itzykson, J.~Zinn-Justin and J.~B.~Zuber,
  Phys.\ Lett.\  B {\bf 82} (1979) 442.

\bibitem{Bena:2003wd}
  I.~Bena, J.~Polchinski and R.~Roiban,
  Phys.\ Rev.\  D {\bf 69} (2004) 046002
  [arXiv:hep-th/0305116].


\bibitem{deVega:1991rc}
  H.~J.~de Vega and F.~Woynarovich,
  J.\ Phys.\ A  {\bf 25}, 4499 (1992).

\bibitem{Aladim:1993uu}
  S.~R.~Aladim and M.~J.~Martins,
  J.\ Phys.\ A  {\bf 26} (1993) 7287
  [arXiv:hep-th/9306049];
  M.~J.~Martins, "Integrable Mixed Vertex Models from Braid-Monoid Algebra",
  [arXiv:solv-int/9903006].

\bibitem{Ribeiro:2005kn}
  G.~A.~P.~Ribeiro and M.~J.~Martins,
  Nucl.\ Phys.\  B {\bf 738} (2006) 391
  [arXiv:nlin/0512035].

\bibitem{abadrios95}
J.~Abad and M.~Rios,
J. Phys. A {\bf 30} (1997) 5887
[arXiv:cond-mat/9706136]; J. Phys. A {\bf 31} (1998) 2269
[arXiv:cond-mat/9801129].

\bibitem{Ragoucy:2007kg}
  E.~Ragoucy and G.~Satta,
  JHEP {\bf 0709}, 001 (2007)
  [arXiv:0706.3327 [hep-th]].

\bibitem{Kulish:1983rd}
  P.~P.~Kulish and N.~Y.~Reshetikhin,
  J.\ Phys.\ A  {\bf 16} (1983) L591.

\bibitem{Gaiotto:2008cg}
  D.~Gaiotto, S.~Giombi and X.~Yin,
  ``Spin Chains in N=6 Superconformal Chern-Simons-Matter Theory,''
  arXiv:0806.4589 [hep-th].

\bibitem{Grignani:2008is}
  G.~Grignani, T.~Harmark and M.~Orselli,
  ``The SU(2) x SU(2) sector in the string dual of N=6 superconformal
  Chern-Simons theory,''
  arXiv:0806.4959 [hep-th].

\bibitem{Minahan:2008hf}
  J.~A.~Minahan and K.~Zarembo,
  ``The Bethe ansatz for superconformal Chern-Simons,''
  arXiv:0806.3951 [hep-th].

\bibitem{Arutyunov:2008if}
  G.~Arutyunov and S.~Frolov,
  ``Superstrings on $AdS_4 x CP^3$ as a Coset Sigma-model,''
  arXiv:0806.4940 [hep-th].

\bibitem{Stefanski:2008ik}
  B.~J.~Stefanski,
  ``Green-Schwarz action for Type IIA strings on $AdS_4\times CP^3$,''
  arXiv:0806.4948 [hep-th].

\bibitem{Gromov:2008bz}
  N.~Gromov and P.~Vieira,
  ``The AdS4/CFT3 algebraic curve,''
  arXiv:0807.0437 [hep-th].


\bibitem{osp-auto}
V.~Serganova, Math. USSR-Izv {\bf 24} (1985) 539; D.~Grantcharov and A.~Pianzola, Int. Math. Res. Not. (IMRN) {\bf 73} (2004) 3937.

\bibitem{Berkovits:1999zq}
  N.~Berkovits, M.~Bershadsky, T.~Hauer, S.~Zhukov and B.~Zwiebach,
  Nucl.\ Phys.\  B {\bf 567} (2000) 61
  [arXiv:hep-th/9907200].

\bibitem{Goldschmidt:1980wq}
  Y.~Y.~Goldschmidt and E.~Witten,
  Phys.\ Lett.\  B {\bf 91} (1980) 392.

\bibitem{Abdalla:1982yd}
  E.~Abdalla, M.~Forger and M.~Gomes,
  Nucl.\ Phys.\  B {\bf 210} (1982) 181.

\bibitem{Bender:1998ke}
  C.~M.~Bender and S.~Boettcher,
  Phys.\ Rev.\ Lett.\  {\bf 80} (1998) 5243
  [arXiv:physics/9712001].

\bibitem{Benna:2008zy}
  M.~Benna, I.~Klebanov, T.~Klose and M.~Smedback,
  ``Superconformal Chern-Simons Theories and $AdS_4/CFT_3$ Correspondence,''
  arXiv:0806.1519 [hep-th].

\bibitem{Chen:1992ee}
  W.~Chen, G.~W.~Semenoff and Y.~S.~Wu,
  Phys.\ Rev.\  D {\bf 46} (1992) 5521
  [arXiv:hep-th/9209005].

\bibitem{Sieg:2005kd}
  C.~Sieg and A.~Torrielli,
  Nucl.\ Phys.\  B {\bf 723} (2005) 3
  [arXiv:hep-th/0505071].

\bibitem{SchaferNameki:2005is}
  S.~Schafer-Nameki and M.~Zamaklar,
  JHEP {\bf 0510} (2005) 044
  [arXiv:hep-th/0509096].

\bibitem{Ambjorn:2005wa}
  J.~Ambjorn, R.~A.~Janik and C.~Kristjansen,
  Nucl.\ Phys.\  B {\bf 736} (2006) 288
  [arXiv:hep-th/0510171].

\bibitem{Arutyunov:2006gs}
  G.~Arutyunov, S.~Frolov and M.~Zamaklar,
  Nucl.\ Phys.\  B {\bf 778} (2007) 1
  [arXiv:hep-th/0606126].

\bibitem{Kotikov:2007cy}
  A.~V.~Kotikov, L.~N.~Lipatov, A.~Rej, M.~Staudacher and V.~N.~Velizhanin,
  J.\ Stat.\ Mech.\  {\bf 0710} (2007) P10003
  [arXiv:0704.3586 [hep-th]].

\bibitem{Janik:2007wt}
  R.~A.~Janik and T.~Lukowski,
  Phys.\ Rev.\  D {\bf 76} (2007) 126008
  [arXiv:0708.2208 [hep-th]].

\bibitem{Fiamberti:2007rj}
  F.~Fiamberti, A.~Santambrogio, C.~Sieg and D.~Zanon,
  ``Wrapping at four loops in N=4 SYM,''
  arXiv:0712.3522 [hep-th].

\bibitem{Keeler:2008ce}
  C.~A.~Keeler and N.~Mann,
  ``Wrapping Interactions and the Konishi Operator,''
  arXiv:0801.1661 [hep-th].

\bibitem{Penedones:2008rv}
  J.~Penedones and P.~Vieira,
  ``Toy models for wrapping effects,''
  arXiv:0806.1047 [hep-th].

\bibitem{Fiamberti:2008sh}
  F.~Fiamberti, A.~Santambrogio, C.~Sieg and D.~Zanon,
  ``Anomalous dimension with wrapping at four loops in N=4 SYM,''
  arXiv:0806.2095 [hep-th].


\bibitem{Nishioka:2008gz}
  T.~Nishioka and T.~Takayanagi,
  ``On Type IIA Penrose Limit and N=6 Chern-Simons Theories,''
  arXiv:0806.3391 [hep-th].

\bibitem{Gromov:2008qe}
  N.~Gromov and P.~Vieira,
  ``The all loop AdS4/CFT3 Bethe ansatz,''
  arXiv:0807.0777 [hep-th].

\bibitem{Grignani:2008te}
  G.~Grignani, T.~Harmark, M.~Orselli and G.~W.~Semenoff,
  ``Finite size Giant Magnons in the string dual of N=6 superconformal
  Chern-Simons theory,''
  arXiv:0807.0205 [hep-th].

\bibitem{dasilva}
  V.~S.~Alves, M.~Gomes, S.~L.~V.~Pinheiro and A.~J.~da Silva,
  Phys.\ Rev.\  D {\bf 61}, 065003 (2000)
  [arXiv:hep-th/0001221].

\bibitem{dasilva2}
  A.~G.~Dias, M.~Gomes and A.~J.~da Silva,
  Phys.\ Rev.\  D {\bf 69}, 065011 (2004)
  [arXiv:hep-th/0305043].


\end{thebibliography}
\end{document}